\documentclass[fleqn,usenatbib]{mnras}

\usepackage{newtxtext,newtxmath}

\usepackage[T1]{fontenc}
\usepackage{ae,aecompl}

\usepackage{graphicx}	
\usepackage{amsmath}	
\usepackage{amssymb}	

\title[Activity and differential rotation in KIC 11560447]{Recurrent starspot activity and  differential rotation in KIC 11560447}

\author[\"{O}zavc{\i} et al.]{
\.{I}. \"{O}zavc{\i}$^{1,}$\thanks{Email: iozavci@ankara.edu.tr}
H.V. \c{S}enavc{\i},$^1$
E. I\c{s}{\i}k,$^{2,3}$
G.A.J. Hussain,$^{4,5}$
D. O'Neal,$^6$
M. Y{\i}lmaz,$^1$
\newauthor{
and
S.O. Selam$^1$}
\\
$^{1}$Department of Astronomy and Space Sciences, Faculty of Science, Ankara University, Tando\u{g}an 06100 Ankara, Turkey\\
$^{2}$Max-Planck-Institut f\"{u}r Sonnensystemforschung, Justus-von-Liebig-Weg 3, 37077, G\"{o}ttingen, Germany\\
$^{3}$Feza G\"ursey Center for Physics and Mathematics, Bo\u{g}azi\c{c}i University, Kuleli/\"Usk\"udar, 34684, Istanbul, Turkey \\
$^{4}$European Southern Observatory, Karl-Schwarzschild-Str. 2, 85748 Garching bei M\"unchen, Germany\\
$^{5}$Universit\'{e} de Toulouse, UPS-OMP, IRAP, 14 Avenue E. Belin, Toulouse, F-31400, France\\
$^{6}$Keystone College, School of Arts and Sciences, La Plume, PA, 18440, USA\\
}

\date{Accepted XXX. Received YYY; in original form ZZZ}

\pubyear{2017}

\begin{document}
\label{firstpage}
\pagerange{\pageref{firstpage}--\pageref{lastpage}}
\maketitle

\begin{abstract}
We present a detailed analysis of surface inhomogeneities on the K1-type sub-giant component of the rapidly rotating eclipsing binary KIC 11560447, using high-precision \textit{Kepler} light curves spanning nearly four years, corresponding to about 2800 orbital revolutions. We precisely determine the system parameters using high-resolution spectra from the 2.1-m Otto Struve Telescope at the McDonald Observatory. We apply the maximum entropy method to reconstruct the relative longitudinal spot occupancy. Our numerical tests show that the procedure can recover large-scale random distributions of individually unresolved spots, and can track the phase migration of up to three major spot clusters. By determining the drift rates of various spotted regions in orbital longitude, we suggest a way to constrain surface differential rotation and show that the results are consistent with periodograms. The K1IV star exhibits two mildly preferred longitudes of emergence, indications of solar-like differential rotation, and a 0.5 to 1.3-year recurrence period in starspot emergence, accompanied by a secular increase in the axisymmetric component of spot occupancy. 
\end{abstract}





\section{Introduction}

Over the last two decades, improvements in observational tools and techniques have enabled us to move forward in understanding the nature of stellar magnetic activity. Doppler imaging and high-precision photometric studies have allowed us to obtain more insight into the overall distribution and evolution of stellar active regions, as well as activity cycles \citep{strassmeier09,kovari14}. 

Revealing surface inhomogeneities and differential rotation patterns on close binary components is important to better understand the structure and dynamics of strong magnetic activity on cool stars. One advantage of investigating active close binaries is that near-synchronous rotation provides a rather firm reference frame.
This makes these targets particularly attractive for studies of differential rotation. Furthermore, in eclipsing binaries, the (orbital) inclination angle can be determined much more precisely than in single stars or non-eclipsing binary systems, leading to precise determination of physical parameters of both components.

Despite the various outcomes, decades of ground-based monitoring of active cool stars have suffered unavoidable temporal gaps ranging from hours to months. Nearly four years of almost uninterrupted high-precision photometry by the \emph{Kepler} space telescope has provided invaluable data for about 150 thousand stars. It has led to a substantial leap in both photometric precision and temporal resolution, making it possible to track one-dimensional photometric signatures of prominent surface features, as well as exoplanet transits. 

Surface differential rotation (hereafter SDR) on cool stars \citep[][and the references therein]{korhonen12,czesla13} is likely to play a significant role in the generation and transport of stellar magnetic fields \citep{isik07, isik11, karak14}, as is the case for the Sun \citep{cs15}. SDR can be estimated via several techniques such as time series Doppler imaging and photometry. 
To quantify SDR, the difference in the rotation rate between the equator and the pole can be defined as 
\begin{equation}
\Delta \Omega := |\Omega_e - \Omega_p| = \alpha\Omega_e.
\label{eq:sdr}
\end{equation}
Here, $\alpha$ measures the amount of relative shear, with $\alpha = 0$ corresponding to rigid-body rotation. The Sun has $\alpha\sim 0.2$. The SDR pattern is categorised as \emph{solar-like} when $\alpha$ > 0, and \emph{anti-solar} when angular velocity increases with latitude ($\alpha$ < 0). The most straightforward way to obtain $\alpha$ requires information on the latitudinal distribution of starspots. Currently, the most reliable method in obtaining the differential rotation profile is the Doppler imaging (DI) technique, which uses high-resolution intensity spectra to map surface brightness distributions \citep{Vogt1983,hatzes98,barnes00,marsden05,ozdarcan16,kovari16}. Zeeman-Doppler Imaging \citep[ZDI;][]{Semel1989} uses high-resolution circularly polarised spectra to map the large-scale surface magnetic fields. It yields higher shear values than DI, presumably because magnetic fields track across a wider range of latitudes than dark spot groups \citep{dunstone08}. 

Photometric spot modelling can also provide estimates for SDR, by  disentangling various rotation periods pertaining to spot groups which are presumably located at various latitudes \citep{mg03,balona16,lehtinen16}. Both photometric and DI methods provide lower limits for $\Delta\Omega$, to the extent of the latitudinal range occupied by spots. For instance, if there are no spots within a latitudinal range around the equator, or if there is a roughly axisymmetric polar cap, the actual strength of differential rotation will be underestimated, because there are no features to measure the periods or phase shifts in those latitudes. 

The accuracy of stellar differential rotation measurements using photometry is {on} average very low, as demonstrated in extensive simulations of several teams by \citet{aigrain15}, which show very low correlations between the simulated and recovered surface shear rates, regardless of the applied method.
\citet{ra15} proposed a new method to detect the sign of SDR (i.e., low (solar-like) or high (anti-solar) latitudes rotating faster), by analysing high-precision long-term \emph{Kepler} light curves via Lomb-Scargle (LS) periodogram analysis. Obtaining false-positive rates below 12 per cent in simulations of spotted stars, they applied the method to 50 single \emph{Kepler} stars and found 5 - 10 single G stars possibly exhibiting anti-solar SDR, while 21 - 34 of them showed solar SDR.

Photometric mapping of stellar surfaces is strongly limited by the one-dimensional nature of the problem and the associated issues with uniqueness and photometric precision \citep{eker99,lanza16}. However, it provides a strong diagnostic in monitoring surface inhomogeneities of magnetically active stars, especially with space-borne photometry and in the case of eclipsing binaries. In spite of the possibility of false detections of `active longitudes' in close binary stars \citep{Jeffers2005}, we demonstrate through numerical simulations that the maximum entropy method (MEM) can provide accurate reconstructions of latitudinally averaged relative spot filling factor, in particular when the stellar parameters are precisely given, in our case thanks to a simultaneous spectro-photometric solution.

Using \textit{Kepler} data together with the determination of system parameters with high-precision spectroscopy, we present results of light curve inversions of the K1IV primary component of the eclipsing binary KIC 11560447, using the {\tt DoTS} code \citep{Cameron1997}. We obtain a measure for the relative spot filling factor as a function of longitude and time with very high temporal resolution, covering $\sim$ 4 years.
Recently, \citet{hackman13} have estimated a lower limit for differential rotation on the single rapidly rotating giant FK Com using ground-based photometry of about $4^\circ/d$, using the phase shifts of the main active longitude and $4.6^\circ/d$ from the observed rotational periods of the active longitude zone. 
With a similar reasoning, we provide a simple method to estimate the differential rotation of spot groups at different latitudes, which involves measuring the phase shifts of individual streaking spot patterns on the K1 sub-giant (Sect.~\ref{sssec:drtime}). The method can provide an alternative estimate for the strength of surface differential rotation, particularly in case of high-precision, uninterrupted photometry. We check the consistency of the results by comparing with seasonal periodograms, as well as the full periodogram. We find an indication for solar-like differential rotation, using the method proposed by \citet{ra15}. We show in Sect.~\ref{sssec:drfreq} how comparative measures of spot activity indicate the presence of a 1.3-year starspot cycle.  

Although photometry alone has never been taken as sufficiently reliable in pinning down SDR and latitudinal spot distributions, we demonstrate the great potential of light curve inversions using high-cadence photometry in providing clues about activity and SDR on active cool stars. 

\subsection{KIC 11560447} 
\label{system}

KIC 11560447 (TYC 3564-1688-1, $V_{max}$=11.1) was classified as an eclipsing binary of the $\beta$ Lyr type by \citet{Prsa2011}. The variability type of the system was also determined by \citet{Slawson2011} as semi-detached $\beta$ Lyr type, with the help of a neural network analysis of the phased light curves. \citet{Baran2011} pointed out that KIC 11560447 is a Double Periodic Variable (DPV), by considering the light curve characteristics of the system. They also estimated the spectral type of the system as K1IV, using the low resolution spectroscopic data (R=850) obtained at the Nordic Optical Telescope (NOT). The effective temperature for the system was given as 4969~K by \citet{Prsa2011}. \citet{Balona2015} noticed that the system includes a flare star having a flare frequency of 0.179 flares/day, observed in short-cadence mode of the \textit{Kepler} data. The light elements of the system are given by \citet{kirk2016} as

\begin{equation}
{\rm BJD}_{\rm MIN1} = 2454953.7303(1) + 0^{\rm d}.5276790(4) \times E.
\label{ligel}
\end{equation}
In Sect.~\ref{sec:analysis}, we further constrain the physical parameters of the binary components. 

\section{Observations and data reduction}

\subsection{The \textit{Kepler} data}
\begin{figure*}
\begin{center}
\includegraphics[width=40pc, height=15pc]{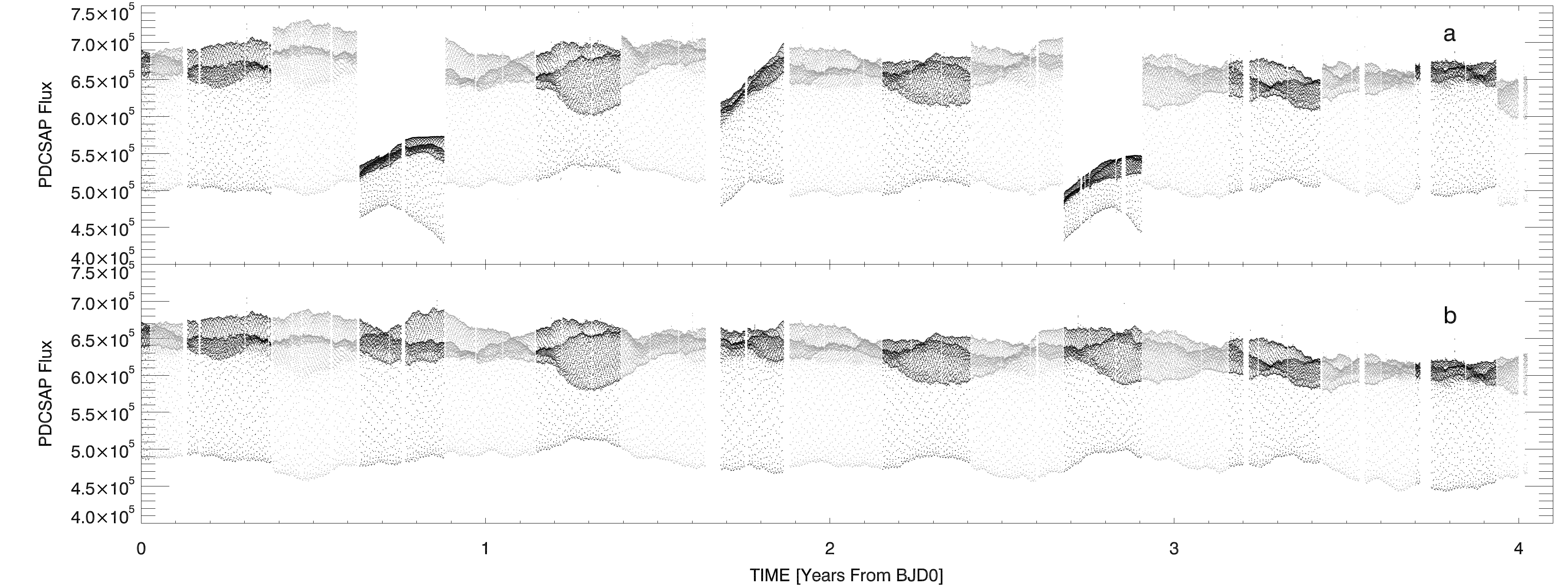}
\caption{Long-cadence (LC) light curves of KIC 11560447 before (a) and after (b) the corrections including photometry with new pixel masks as well as the stitching process.}
\label{fig:stitch}
\end{center}
\end{figure*}

\begin{figure*}
\begin{center}
\includegraphics[width=30pc, height=10pc]{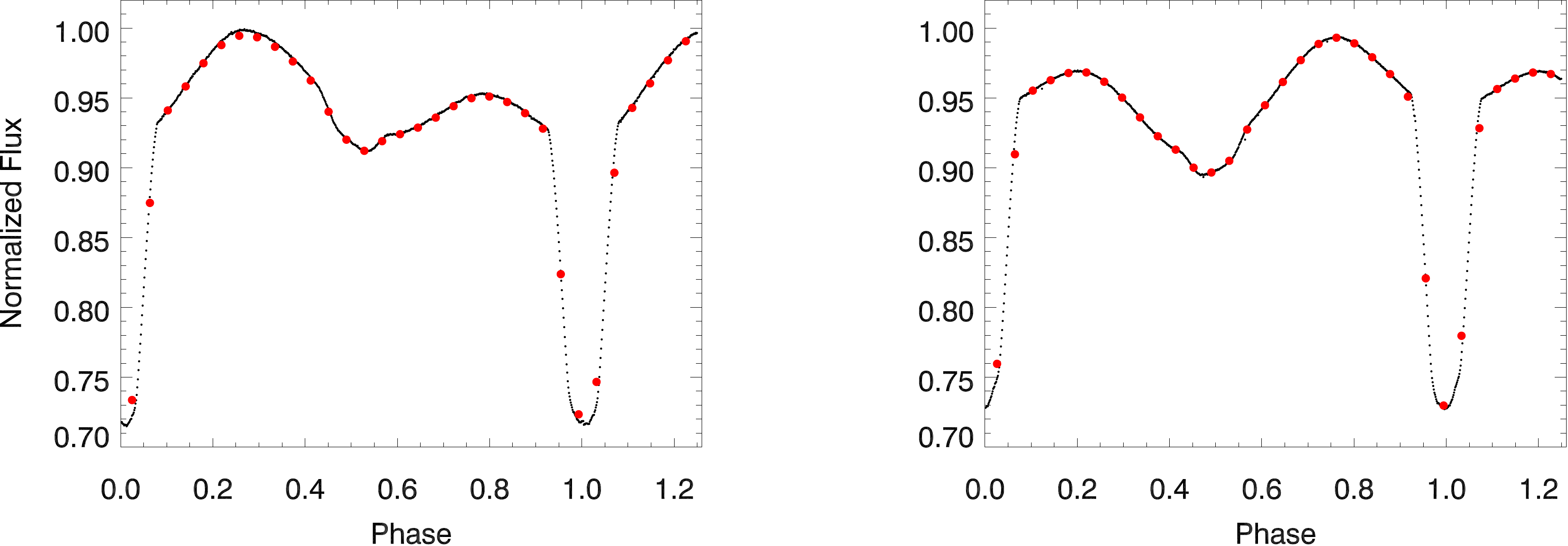}
\caption{Sample light curves of KIC 11560447 from $2^{\rm nd}$ and $12^{\rm th}$ quarters. Black dots represent the SC data, while the superimposed red circles belong to LC data of the corresponding orbital cycles.}
\label{lcsc}
\end{center}
\end{figure*}

The \textit{Kepler} data of the system include 18 quarters (0-17) for Long-Cadence (hereafter LC) and 4 quarters (1, 2, 12, 13) for Short-Cadence (hereafter SC) data. Each orbital cycle estimated using equation~\ref{ligel} includes about 775 and 26 data points for SC and LC data, respectively.  During the analysis, we used all the \textit{KEPLER} PDC (pre-search data conditioning) light curves of the system that are available at the Multimission Archive at STSci (MAST) database. We use PyKE software \citep{Still2012}, in order to check if there is any flux contamination in the target light curve by considering the target pixel files for each quarter using the PyKE task \textit{keppixseries}. During those inspections, we encountered PDC-pipeline specific misinterpretations in the autonomous determination of the target pixel files, especially in quarters \#4, \#8 and \#12, as can  be seen clearly in Figure~\ref{fig:stitch}a. The solution for this ambiguity is found by copying the pixel mask used in quarter \#16 and re-performing photometry for the problematic quarters mentioned above, {this time using the PyKE tasks \textit{kepmask} and \textit{kepextract}}. Finally, the stitching of the data is performed using the approach by \citet{reval2014} and \citet{kasli2015}, by matching the average flux of the last 50 points of the leading quarter to the first 50 points of the trailing quarter. The resulting light curves, for which the flux levels and the amplitudes have been corrected, are shown in Figure~\ref{fig:stitch}b. Two sample light curves, which are based on those corrected data are given in Figure \ref{lcsc}. Both SC and LC time series provide hints about the high level of activity in KIC 11560447.\\

\subsection{Spectroscopy}

The high resolution time-series spectra of KIC 11560447 were obtained using the Sandiford Echelle Spectrograph (SES) attached to the 2.1m Otto Struve Telescope at the McDonald Observatory, in June 2011. We obtained 10 spectra with a mean resolution of $\sim$ 60000 and wavelength coverage from 6490 \r{A} to 8900 \r{A}. We have exposure times ranging between 900 and 2700 seconds, due to bad weather conditions and the low brightness of the target. Signal-to-noise ratios (SNRs) range from <40 (for 900 s exposures) to between 60-80 (1200-2700 s exposures), although 2700 s does also result in phase smearing,  which becomes noticeable for exposures above 5\% of the orbital phase, corresponding to $\sim$ 2280 seconds in our case. The log of the spectroscopic observations is given in Table \ref{speclog}. The reduction procedure such as bias subtraction, flat fielding, removal of cosmic rays, extraction of orders and wavelength calibration were performed using the standard packages of IRAF (Image Reduction and Analysis Facility). Normalisation of the spectra is carried out using a Python code, which performs spline and polynomial fits by considering the related model atmosphere of the target star with the help of the software \textit{iSpec} developed by \citet{Cuaresma2014}.\\

In order to obtain high SNR values for each spectrum, we applied the technique of Least-Squares Deconvolution \citep[LSD; for more details, see][]{Donati1997}. An input line list required for LSD was obtained from the Vienna Atomic Line Database (VALD) \citep{Kupka1999}, according to the $T_{\rm eff}$ and $\log g$ parameters of the system. During the preparation of the line list, several wavelength regions, including strong chromospheric lines (e.g., $\rm H_{\alpha}$) and strong telluric regions, were removed. The SNR values of the input spectra are calculated as the mean square-root of count values from several continuum regions and are between 40-80, while the resultant LSD profiles also have low SNR values in the range 175 - 300. Such low enhancement of SNR even with the LSD technique is due to the few number of lines (only 270) used in the convolution. The reasons are $(i)$ the relatively few photospheric lines in the observed wavelength region (6490 - 8900 \r{A}), and $(ii)$ strong rotational broadening, for which relatively weak lines blend with the continuum at our SNR levels. In addition, the SES spectrograph has a well-known issue of internal reflection, affecting 4 echelle orders (covering $\sim$ 400 \r{A}) with excess illumination. These limitations have thus led to an enhancement in SNRs by a factor of about 4, which allowed us to infer only the radial velocities from the spectra. 

\begin{figure}
\includegraphics[width=20pc]{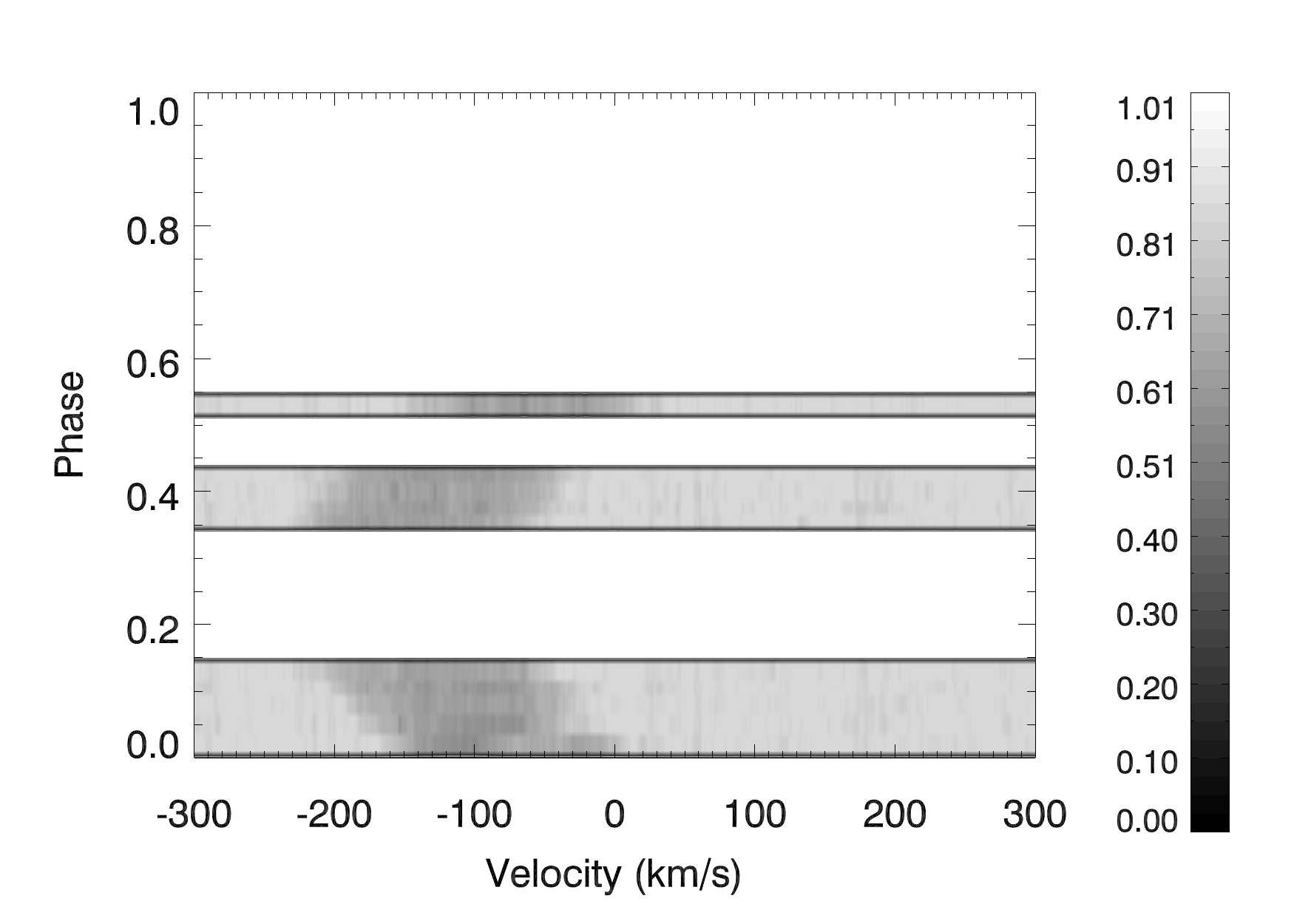}
\caption{{A time-series grey-scale plot representation of the LSD profiles obtained using 10 spectra of KIC 11560447. Note that the intensity contribution of the secondary component is only slightly visible at phase 0.0, as a blended pattern at redder velocities of the profile.}}
\label{fig:lsd}
\end{figure}

As the resultant LSD profiles have low SNR values, the profile widths and depths can vary considerably, especially for the ones obtained using 900 secs of exposure time. In addition, the higher SNR data cover a very small fraction of the orbital phases (only the eclipses). This phase coverage was insufficient to reconstruct a reliable spot map using Doppler imaging. The LSD profiles showing the phase coverage are given in Fig.~\ref{fig:lsd}. We have found radial velocity (hereafter RV) data points of the primary component and only one RV data point for the secondary component, corresponding to the phase 0.01 via \textit{iSpec} using the cross-correlation function (CCF), which could then be used during simultaneous light and radial velocity analysis. {As input data for CCFs we used the spectral data covering the wavelength range from 6570 to 6850 \r{A}, except strongly blended lines and broad lines such as $\rm H_{\alpha}$. We used the spectrum of HR 7345 (G8V) as the template spectrum obtained also at the McDonald Observatory.}

\begin{table}
\caption{Log of spectroscopic observations.}
\label{speclog}
\centering
\begin{tabular}{ccccc}
\hline
Date&Exp.&HJD&Phase&{SNR}\\
 & Time(s) &Mid Time\\
\hline
06/09/11	&	2700	&	55722.8325	&	0.52	&	56.3	\\
06/10/11	&	2700	&	55723.8287	&	0.41	&	78.4	\\
06/11/11	&	900		&	55724.8532	&	0.35	&	34.3	\\
06/11/11	&	900		&	55724.8642	&	0.37	&	33.1	\\
06/11/11	&	900		&	55724.8752	&	0.39	&	32.0	\\
06/14/11	&	1200	&	55727.8427	&	0.01	&	59.9	\\
06/14/11	&	1200	&	55727.8571	&	0.04	&	61.4	\\
06/14/11	&	1200	&	55727.8715	&	0.07	&	63.6	\\
06/14/11	&	1200	&	55727.8859	&	0.10	&	64.4	\\
06/14/11	&	1200	&	55727.9002	&	0.12	&	63.8	\\

\hline
\end{tabular}
\end{table}

\section{Analysis}
\label{sec:analysis}

\subsection{System parameters}
\label{ssec:photo}
\begin{figure}
\begin{center}
\includegraphics[width=\columnwidth]{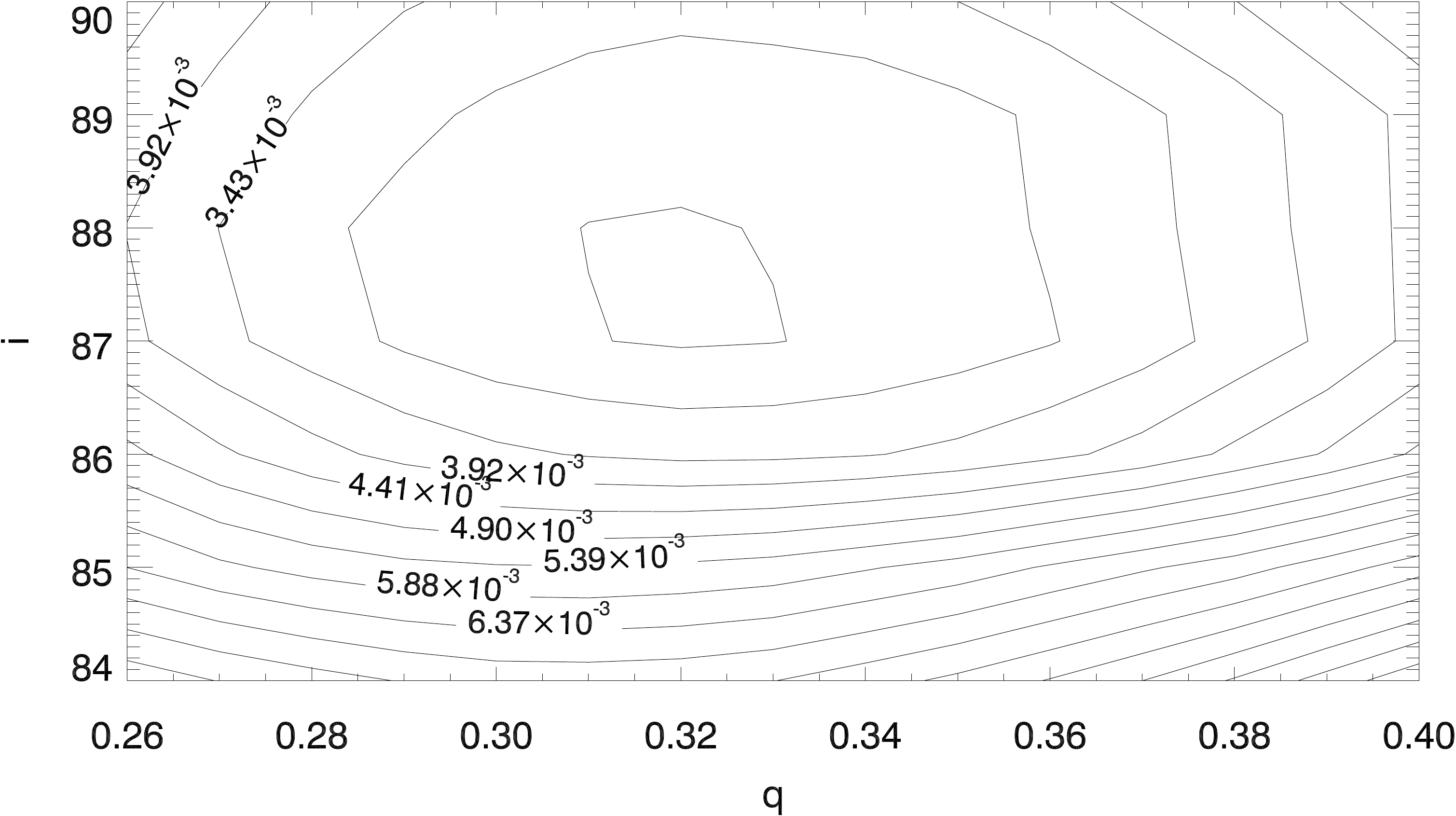}
\caption{Contour plot of 2D $q$-$i$ grid search and their $\chi^2 = \Sigma(O-C)^2$ values using the 30$^{\rm th}$ cycle using SC data. The innermost contour line corresponds to $\chi^2=2.45\times 10^{-3}$.}
\label{qsearch}
\end{center}
\end{figure}

The physical parameters of KIC 11560447 were obtained using the {{\tt PHOEBE} software \citep{Prsa2005}, a graphical user interface that is based on the 2003 version of the Wilson-Devinney code \citep{Wilson1971}}. During the light curve analysis, we have chosen the 30$^{\rm th}$ cycle of the short-cadence data as the `clean' light curve, since the shape of the light curve is almost symmetric as a consequence of minimum rotational modulation of starspot activity. Considering the total eclipse seen in the secondary minimum, we performed a 2-dimensional mass ratio and orbital inclination ($q$-$i$) grid search using the \emph{scripter} mode of {\tt PHOEBE}, with step sizes in $q$ and $i$ as 0.01 and 0.10, respectively. {During the grid search, the dimensionless surface potential of the components ($\Omega_{1,2}$), the luminosity of the primary component ($L_1$) and the effective temperature of the secondary component ($T_2$) are set as free parameters. The surface albedos ($A_{1,2}$) as well as the gravity darkening of both components ($g_{1,2}$) are set to 0.5 and 0.32, respectively, suitable for convective envelopes \citep{vonzeipel24,lucy64,rucinski69}.} As a first attempt, we carried out a grid search assuming a semi-detached model, however it was not possible to find a converged solution. With the detached mode, iterations converged to a solution and gave reasonable results. However, it was not possible to fit the observed light curve, probably due to the existence of cool starspot(s) on the primary component. Therefore, considering the best-fit $(q, i)$ pair obtained from the spotless grid search, we simply attempted to fit a circular-shaped cool spot on the primary component, this time using the GUI mode of {\tt PHOEBE} by additionally adjusting  \textit{q} and \textit{i} until the minimum $\Sigma(O-C)^2$ has been achieved. The resulting spot parameters are latitude = 63$^{\circ}$, longitude = 7$^{\circ}$, temperature factor = 0.8, and radius = 26$^{\circ}$. Afterwards, using the above spot parameters, we ran another \textit{q-i} grid search and determined the best-fitting \textit{q-i} pair under the spotted approximation. The resulting distribution for {$\chi^2 = \Sigma(O-C)^2$} from the spotted 2-dimensional grid search are shown in Fig.~\ref{qsearch} (the $\chi^2$ values are obtained at the grid points only). This spotted solution gave us a more compatible fit to the observational light curve for ${\rm min}(\chi^2)$, corresponding to $q=0.32$ and $i=87.5$. We note that the results are consistent, whether or not by including the measurement errors into the $\chi^2$, given the small variation in the Kepler errors over the observations.

In addition to the \textit{q-i} grid search, we have also performed simultaneous light and radial velocity curve analysis, using the GUI-mode of {\tt PHOEBE}. The best-fit \textit{q} and \textit{i} values obtained from the 2D grid search are used as preliminary values for the simultaneous solution. During the iterations, albedos ($A_1=A_2$) and gravity darkening coefficients ($g_1=g_2$) of the components were assumed as 0.5 and 0.32, respectively, as performed during \textit{q-i} grid search. The limb darkening coefficients were taken from \citet{claret11}. This analysis gave similar \textit{q} and \textit{i} results to that obtained from the \textit{q-i} grid search.

\begin{figure}
\begin{center}
\includegraphics[width=\columnwidth]{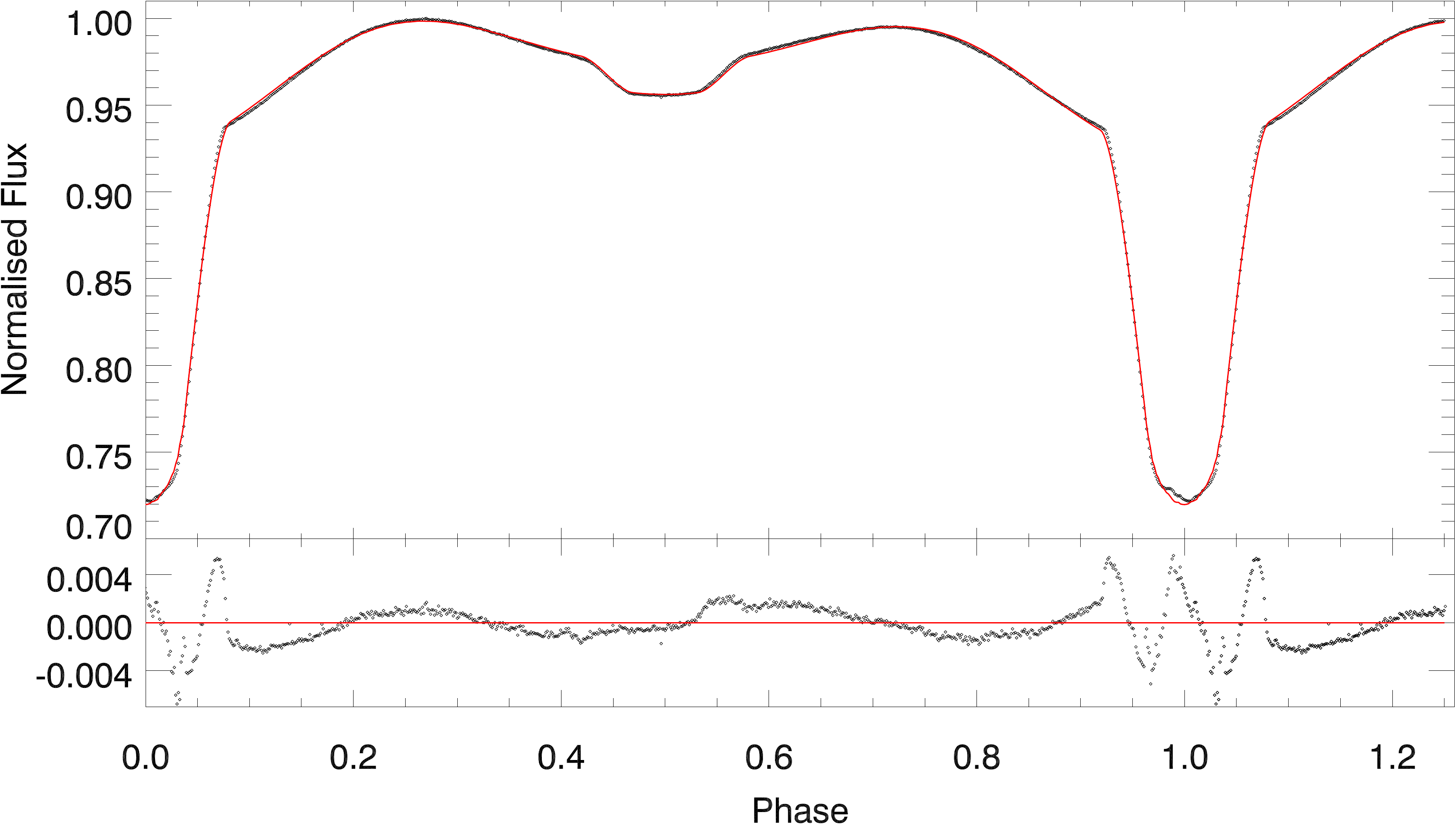}
\caption{\emph{Upper panel}: the observational (black dots) and the model (red solid line) light curve of KIC 11560447, based on the simultaneous light curve and radial velocity data analysis of 30$^{\rm th}$ cycle using SC data. \emph{Lower panel}: the residuals of the data from the model. The model light curve includes a circular-shaped spot with 63$^{\circ}$ latitude, 7$^{\circ}$ longitude, a temperature factor of 0.8, and radius 26$^{\circ}$.}
\label{lcfit}
\end{center}
\end{figure}

\begin{figure}
\begin{center}
\includegraphics[width=1.05\linewidth]{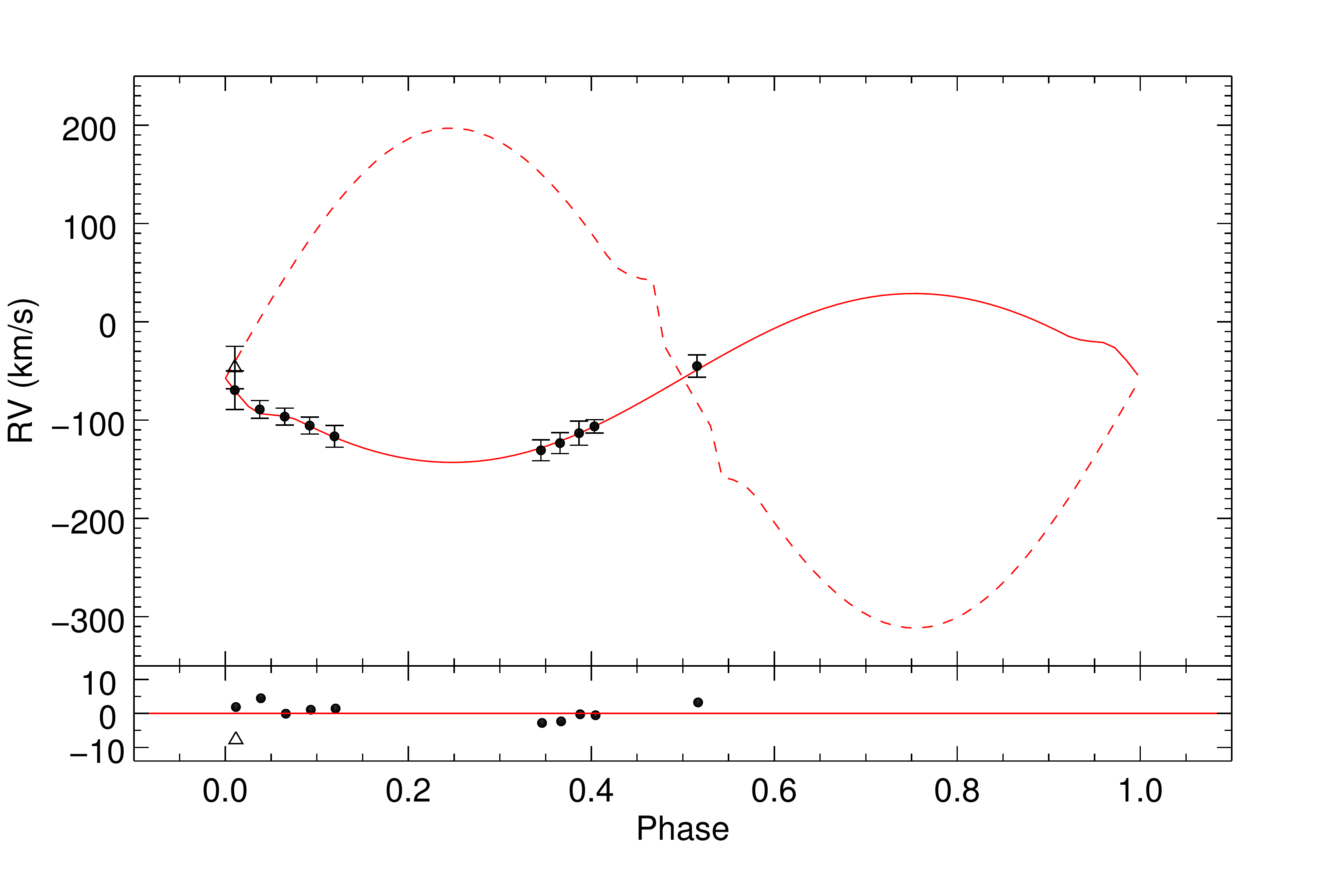}
\caption{The radial velocity (RV) curve of KIC 11560447 (upper panel). The solid and dashed red lines represent the RV fit to the data obtained via {\tt PHOEBE} analysis, whereas the filled circles and the open triangle denote the RV data of the primary and secondary components, respectively. The lower panel shows the residuals of the data from the model.} 
\label{RVs}
\end{center}
\end{figure}

The parameters obtained from simultaneous analysis are given in Table~\ref{table2} {and} used as input parameters during photometric mapping with the {\tt DoTS} code (Sect. ~\ref{spotmodel}-\ref{spotmodelkic}). The best fits to the light curve and RV data are given in Figs.~\ref{lcfit} and \ref{RVs}, respectively.

\begin{table}

\caption{Results from the simultaneous light and radial velocity curve analysis (estimated errors in parentheses are based on differential corrections using the {\tt PHOEBE} code).}
\label{table2}
\centering
\begin{tabular}{ll}
\hline
\renewcommand{\thefootnote}{\fnsymbol{footnote}}
Parameter\footnotemark[1] & Value \\
 
\hline
$T_1$ [K]			&	4969				\\
$T_2$ [K]			&	3068(17)			\\
${L_1/(L_1+L_2)}$ 	&	0.97(1)				\\
$\Omega_1$			&	3.32(1)				\\
$\Omega_2$			&	3.66(1)				\\
$q=M_2/M_1$	&	0.33(1)				\\
$i$ [$^{\circ}$]		&	88.2(1)				\\
$V_\gamma$ [km/s]	&	-57.2(2)			\\
$a$[$R_\odot$]		&	3.59(1)				\\
$R_1$ [$R_\odot$]	&	1.22(1)				\\
$R_2$ [$R_\odot$]	&	0.52(1)				\\
$M_1$ [$M_\odot$]	&	1.68(3)				\\
$M_2$ [$M_\odot$]	&	0.56(1)				\\
$M_{\rm bol, 1}$			&	4.97(1)				\\
$M_{\rm bol, 2}$ 		&	8.94(4)				\\
$\log$ \emph{$g_1$}	&	4.49(1)				\\
$\log$ \emph{$g_1$}	&	4.76(1)				\\
\hline
\end{tabular}
\begin{flushleft} \footnotesize

$^{\star}$ $T_{1,2}$ - temperature of the primary and secondary, ${L_1/(L_1+L_2)}$ - fractional luminosity of the primary, $\Omega_{1,2}$ - dimensionless surface potentials, $q=M{_2}/M{_1}$ - mass ratio of components, $i  [^\circ]$ - orbital inclination, V$_\gamma$ - velocity of the mass centre in km/s, $a~[R_{\odot}]$ - orbital semi-major axis, ${{R}_{1,2} [R_{\odot}]}$, ${{M}_{1,2} [M_{\odot}]}$ - stellar masses and mean radii, $M_{\rm bol1,2}$ bolometric magnitudes, ${\log} \ g{_{1,2}}$ - logarithm of the mean surface gravity (cgs units).
\end{flushleft}
\end{table}

\subsection{Spot modelling: numerical tests} 
\label{spotmodel}

Before the photometric mapping of KIC 11560447 data via {\tt DoTS}, we have carried out numerical simulations to test the reconstruction performance of the light-curve inversion mode. 

\begin{figure}
\includegraphics[width=\linewidth,angle=0]{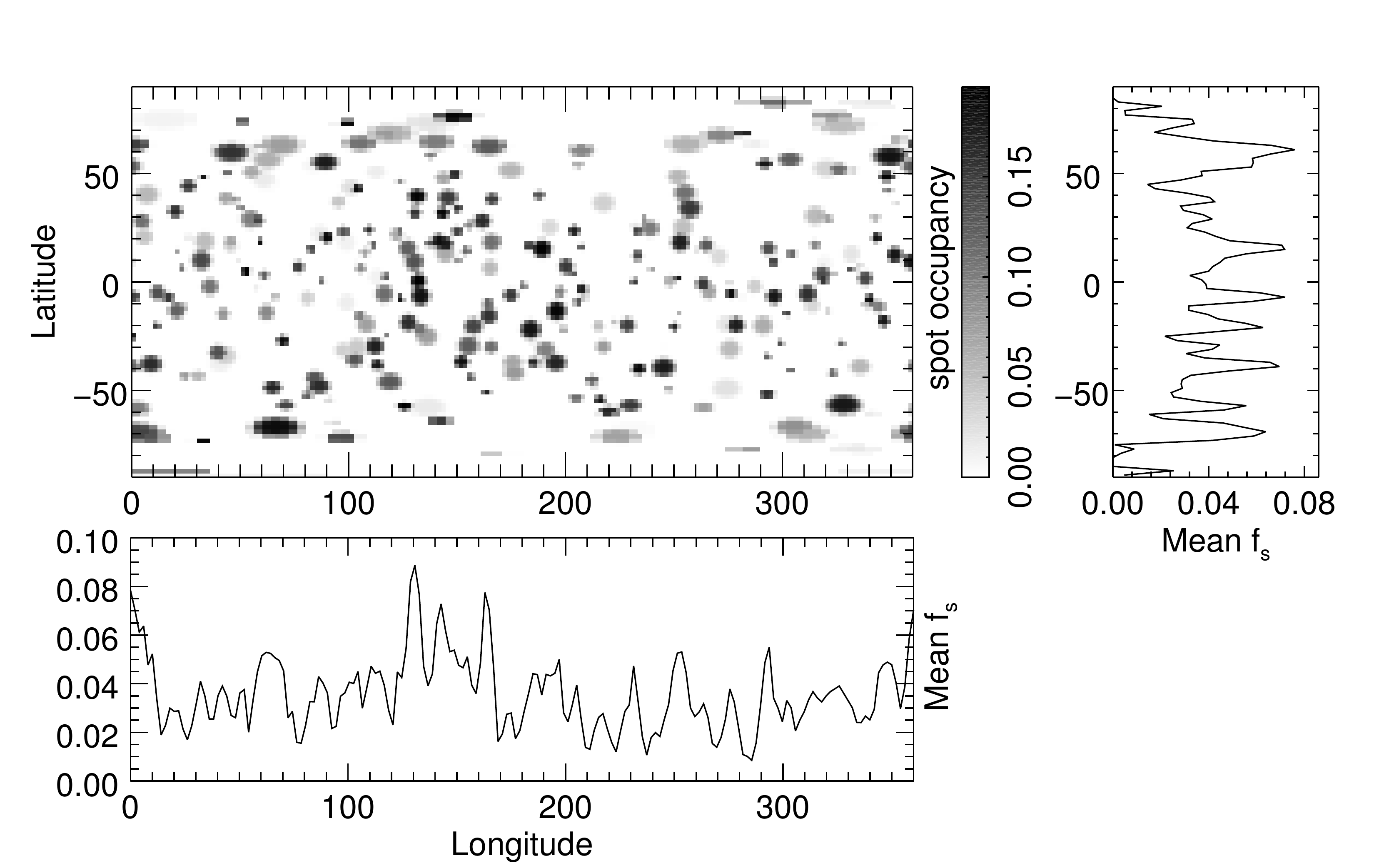}
\includegraphics[width=\linewidth,angle=0]{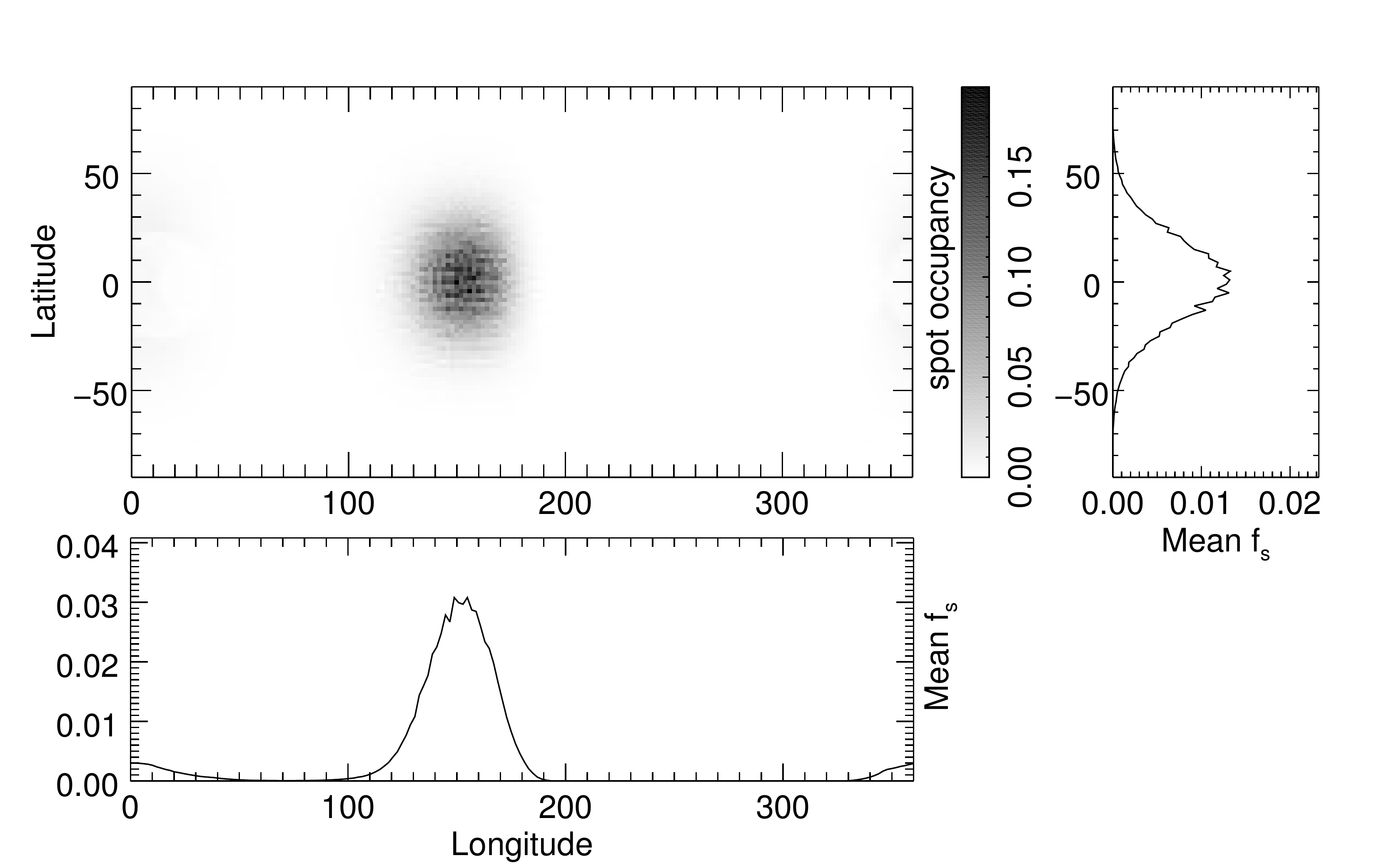}
\caption{Surface maps of spot occupancy and their longitudinal and latitudinal profiles, 
for an example input distribution with random locations (upper panel) and the corresponding {\tt DoTS} reconstruction (lower panel).}
\label{sim}
\end{figure}

In a first test, we placed 300 spots with a uniform area distribution with radii in the range $[1^\circ,4^\circ)$, at uniformly distributed random longitudes $\phi$ and random latitudes $\lambda\in[-\pi/2,\pi/2]$ with a distribution $\lambda=\sin^{-1}(2x+1)$, where $x\in[0,1]$, following \citet{Jeffers2005}. We generated synthetic light curves based on {$10^3$} different random realisations for input maps, using the physical and geometric parameters of the system from Table~\ref{table2}, and the sampling rate equivalent to that of LC data. 
We then used those synthetic light curves for surface reconstructions. To determine the degree of convergence for maximum entropy, we used the {\tt TEST} statistics option of {\tt DoTS}, which enables us to estimate where the procedure begins fitting noise rather than spot signatures (see \cite{senavci11} for more details). An example for the input and reconstructed maps is shown in Fig.~\ref{sim} (in all maps, the longitude runs in reverse to phase). 
The procedure recovers the randomly formed clustering of relatively strong spots around $150^\circ$ longitude reasonably well.

\begin{figure}
\begin{center}
\includegraphics[width=19.6pc]{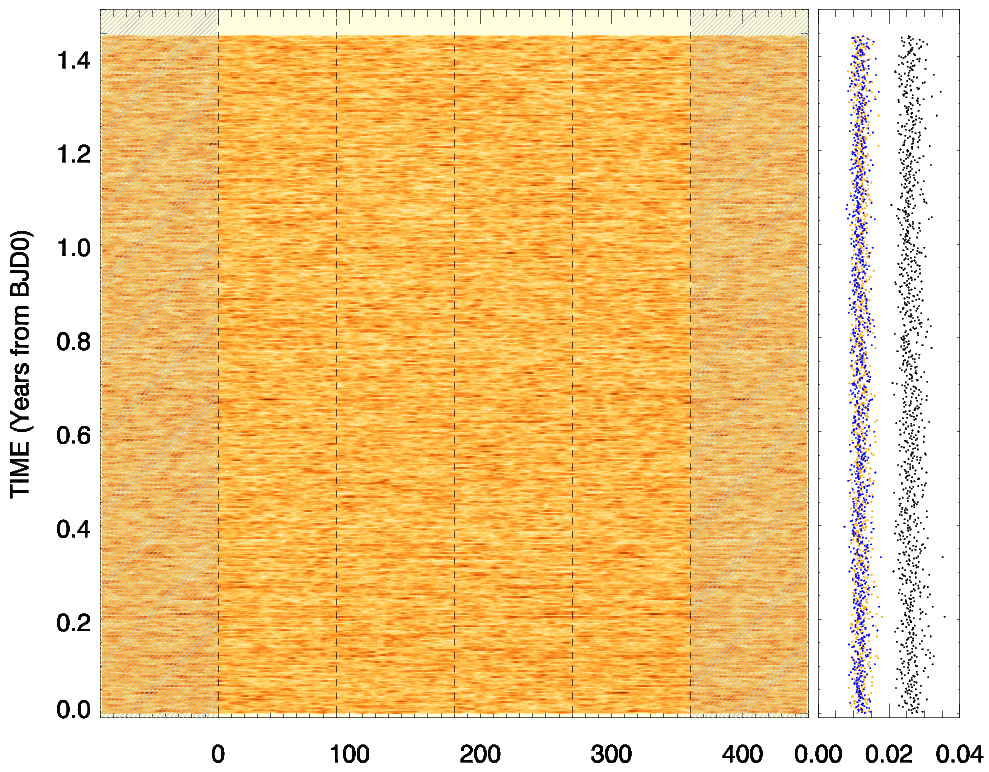}
\end{center}
\vspace{-7.cm} \flushleft (a) \vspace{6.20cm} 
\begin{center}
\vskip-1mm\includegraphics[width=20pc]{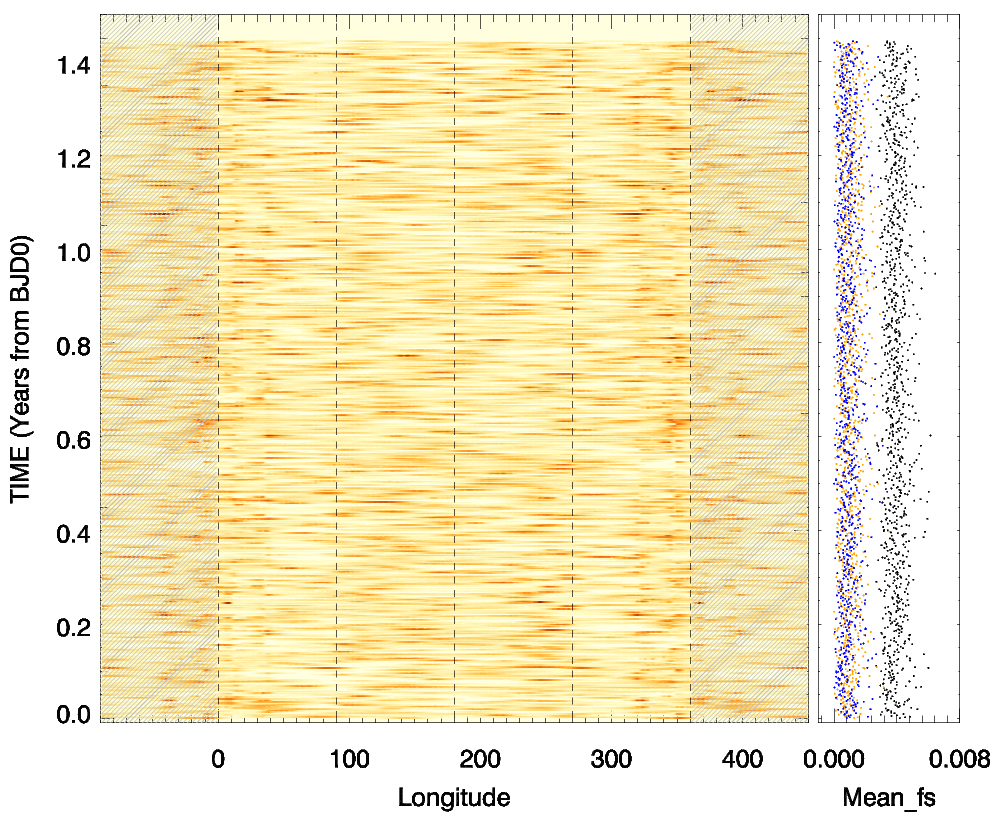}
\end{center}
\vspace{-7.5cm} \flushleft (b) \vspace{6cm} 
\hskip+2.8mm
\flushleft (c)
\begin{center}
\hskip-4mm\includegraphics[width=17.5pc]{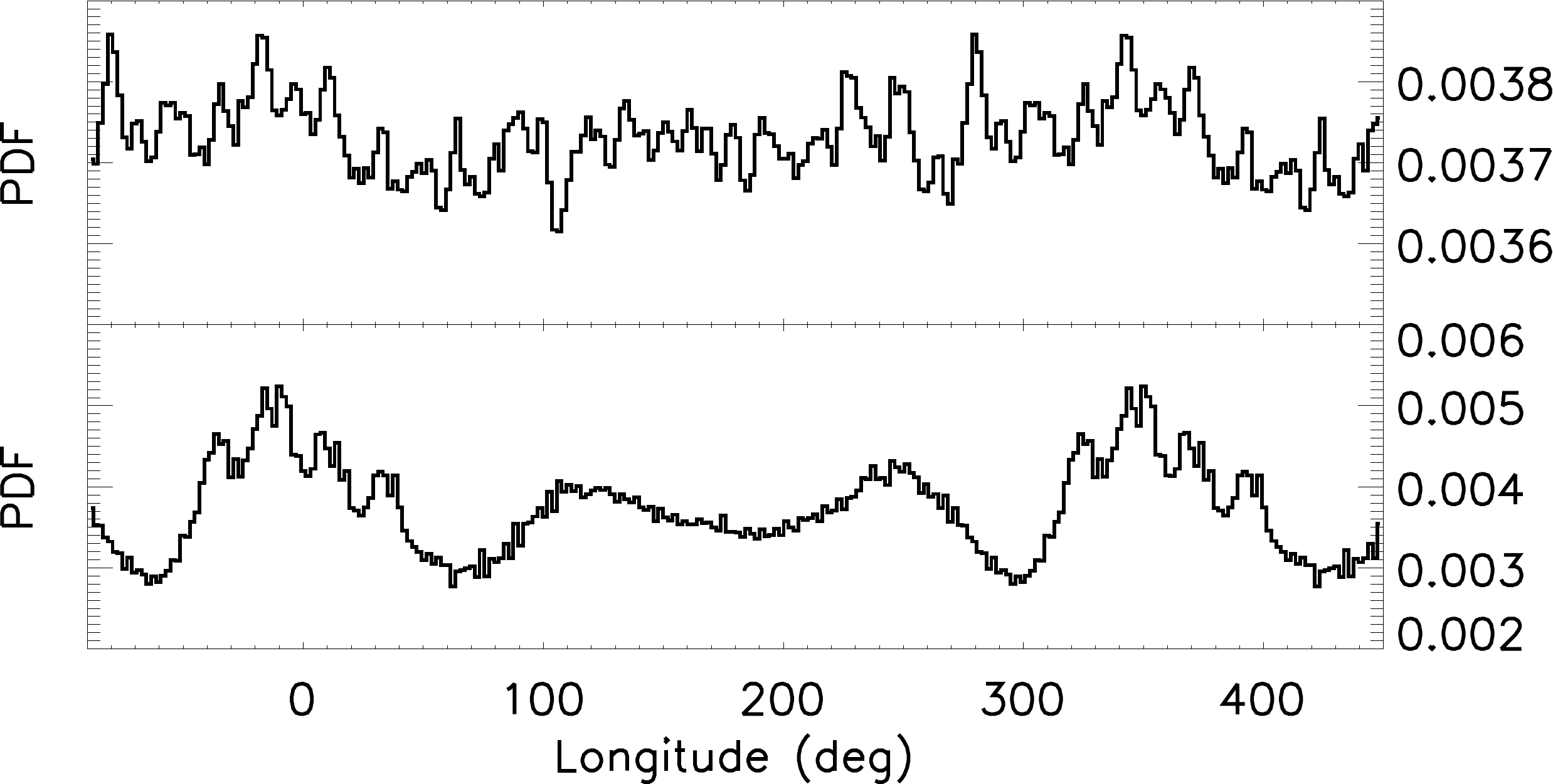}
\end{center}
\vskip-2mm\caption{Input (a) and reconstructed (b) longitudinal 
maps of relative spot occupancy on the K1IV star, where orbital cycles are modelled with random spot distributions. To enhance visibility, the plots in the main panels are extended for a quarter period in each longitudinal direction (shaded regions). The right panels show the mean relative spot occupancy for the visible hemisphere centred at a longitude of 90$^\circ$ (orange), 270$^\circ$ (blue), and their sum (black offset by +0.002). (c) The probability distributions obtained from the above maps (input above, reconstructed below).}
\label{rand}
\end{figure}

Figure~\ref{rand}a and b show, respectively, the input and reconstructed distributions of random spots for the entire set of $10^3$ realisations, in the form of latitudinally averaged mean spot filling factor\footnote{We use the terms spot filling factor, spot occupancy, and spot coverage interchangeably, to point out to the latitudinally averaged \emph{relative} $f_s(\phi,t)$, which is rescaled for each orbital cycle so that zero filling factor corresponds to the maximum brightness in that cycle.} $f_s(\phi,t)$, which is a function of longitude $\phi$ and time $t$. 
The right panels show that there is no systematic hemispheric preference in $\langle f_s\rangle$. 
Intriguingly, {\tt DoTS} can reconstruct randomly forming large-scale spot conglomerates as in Fig.~\ref{sim}, though it fails for individual spots, as expected. This is also seen in Fig.~\ref{rand}c, where we show the probability distribution functions (hereafter PDF) derived from the input and output maps above by integrating over time, over 2-degree longitudinal bins and normalising by the integral over longitude and time. 
The overall distributions show similarity, except for small-scale variations, as expected. Relatively large deviations from the input model are concentrated in the primary eclipse zone ($\phi=0^\circ\pm 30^\circ$), where the time resolution of the LC mode is insufficient (see Fig.~\ref{lcsc}). 
We note that the ranges of the mean $f_s$ and the PDFs in the present and the following cases are different for the input and reconstructed maps.

\begin{figure}
\includegraphics[width=\linewidth,angle=0]{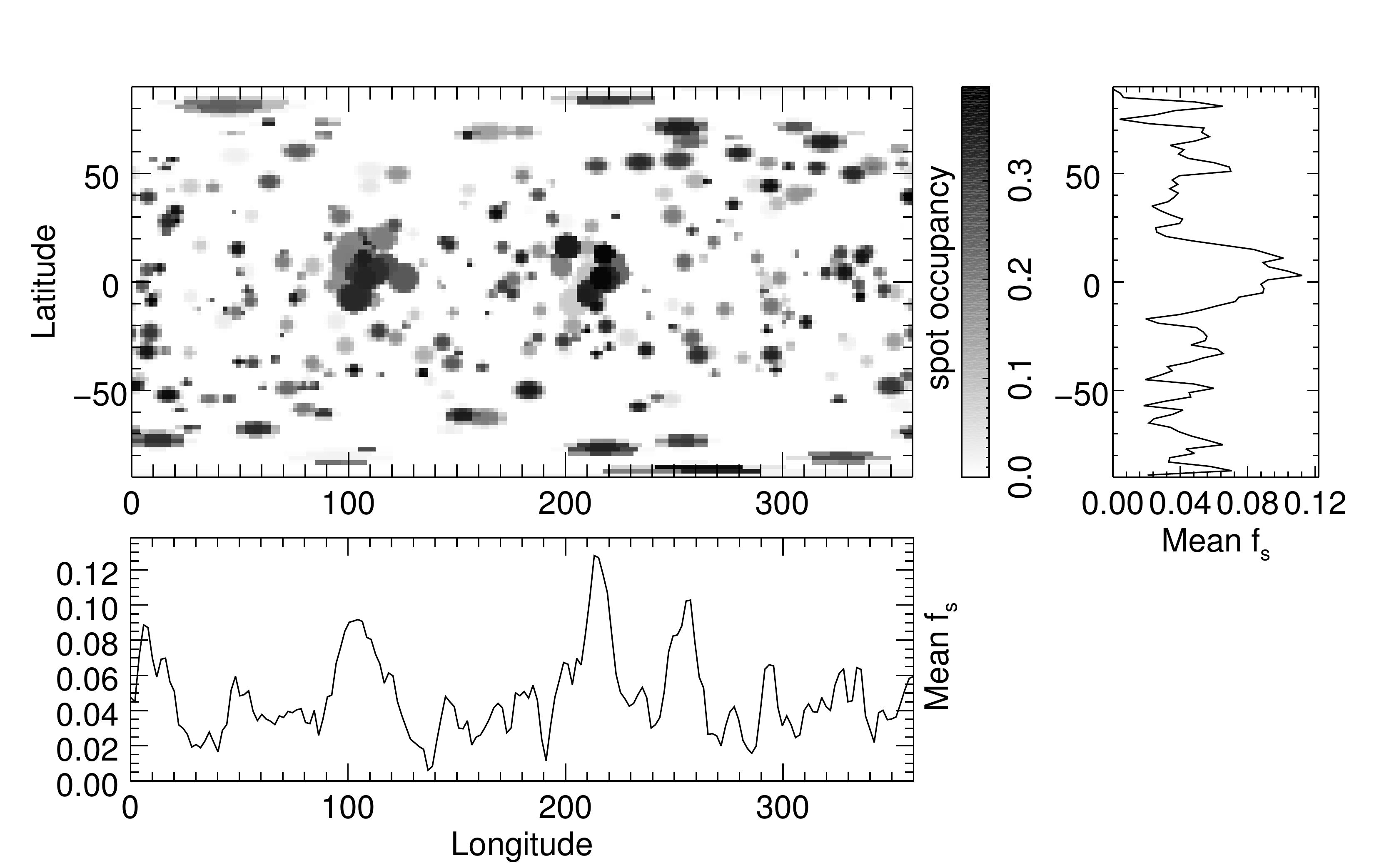}
\includegraphics[width=\linewidth,angle=0]{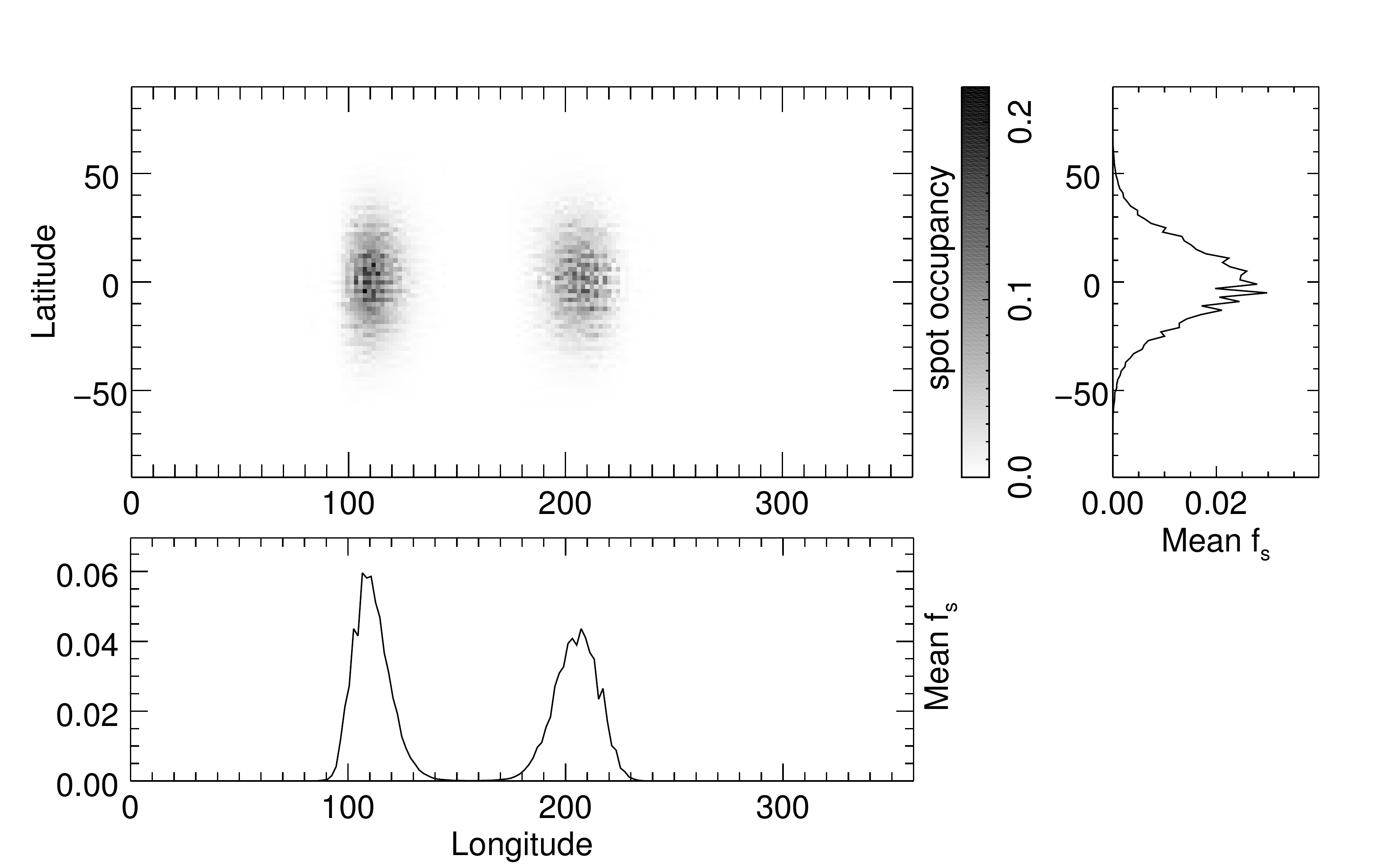}
\caption{{Same as Fig.~\ref{sim}, for two migrating spot groups centred at $110^\circ$ and $210^\circ$ longitudes.}}
\label{sim_mig}
\end{figure}

In a second simulation, we have tested the capability of {\tt DoTS} in detecting longitudinal migration of large-scale spot groups. Initially, we took an input surface map with 300 randomly distributed spots (with the same properties otherwise as the first simulation) and added two large spot groups consisting of 20 spots each, with the same mean contrast level as the background spots, but with radii ranging between $4^\circ$ to $7^\circ$, so that their mean surface area is larger by a factor of about three. The longitudinal separation of the two spot clusters is chosen to be 100$^\circ$ and they are both centred at the equator. They are shifted towards increasing longitudes (prograde) with a rate of $1^\circ$ per orbital revolution for both clusters ($\sim 1.9^\circ/$d), while the spots keep their size and relative locations within each group. The light curve was generated again with the LC sampling rate. An example input map and the corresponding reconstruction are shown in Fig.~\ref{sim_mig}. Here, migrating spot groups are visible around $110^\circ$ and $210^\circ$ longitudes. 

\begin{figure}
\begin{center}
\hskip-2mm\includegraphics[width=19.6pc]{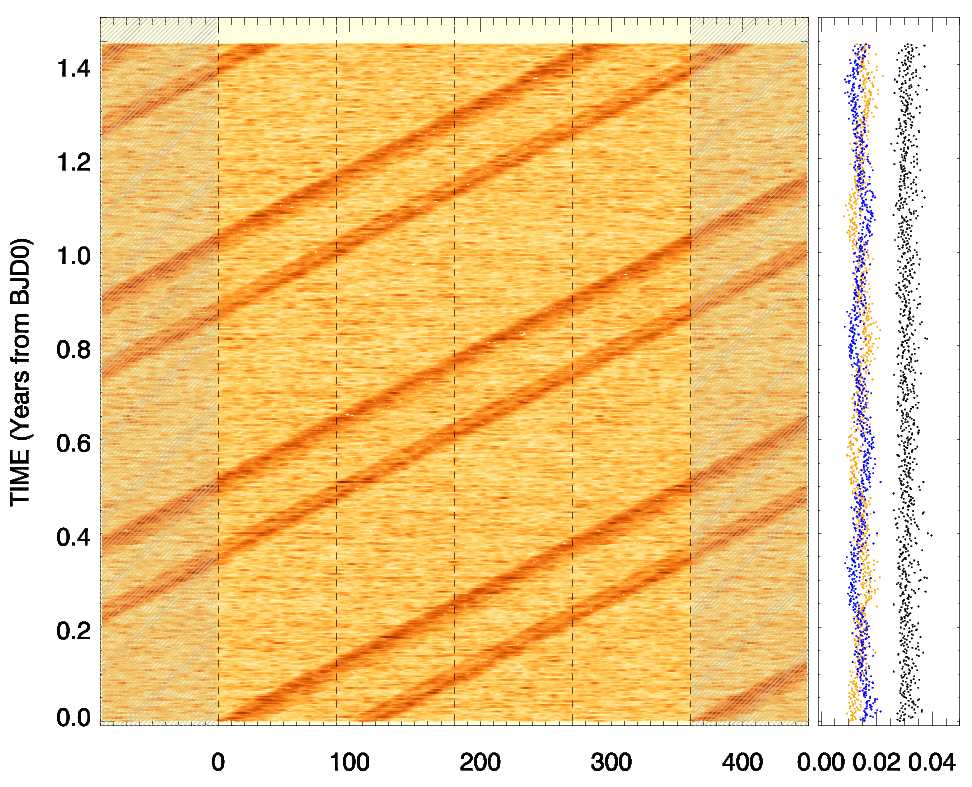}
\end{center}
\vspace{-7.cm} \flushleft (a) \vspace{6.20cm} 
\begin{center}
\vskip-1mm\includegraphics[width=20pc]{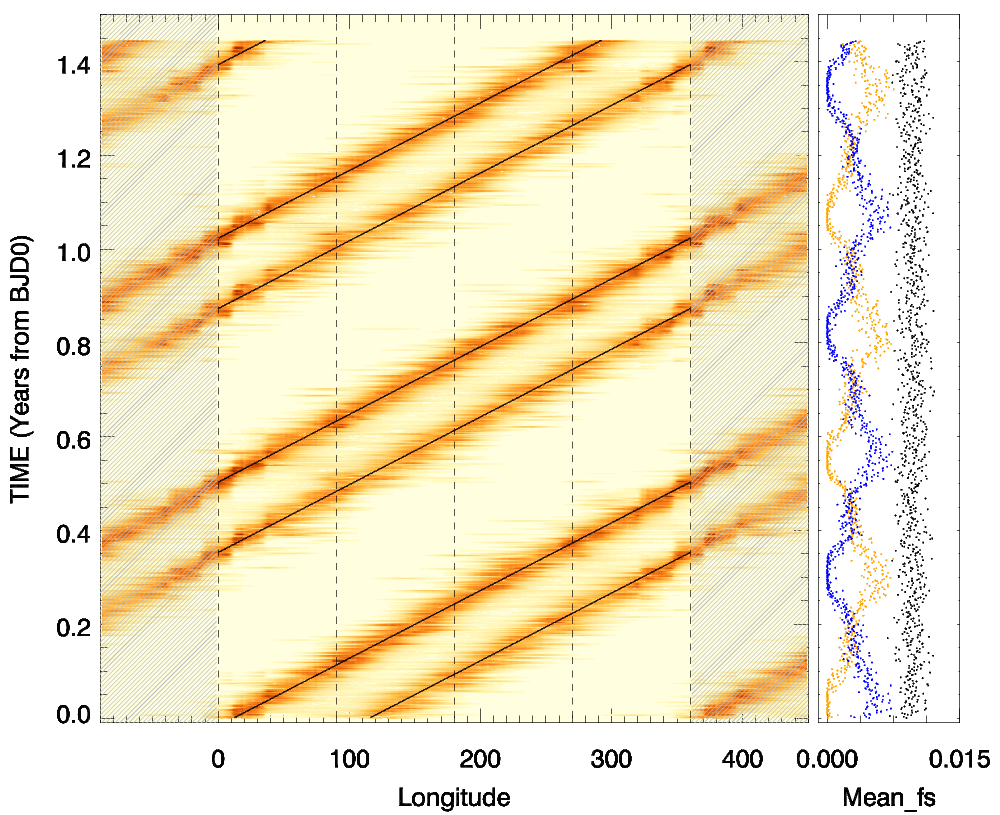}
\end{center}
\vspace{-7.5cm} \flushleft (b) \vspace{6cm} 
\hskip+2.8mm
\flushleft (c)
\begin{center}
\hskip-4mm\includegraphics[width=17.5pc]{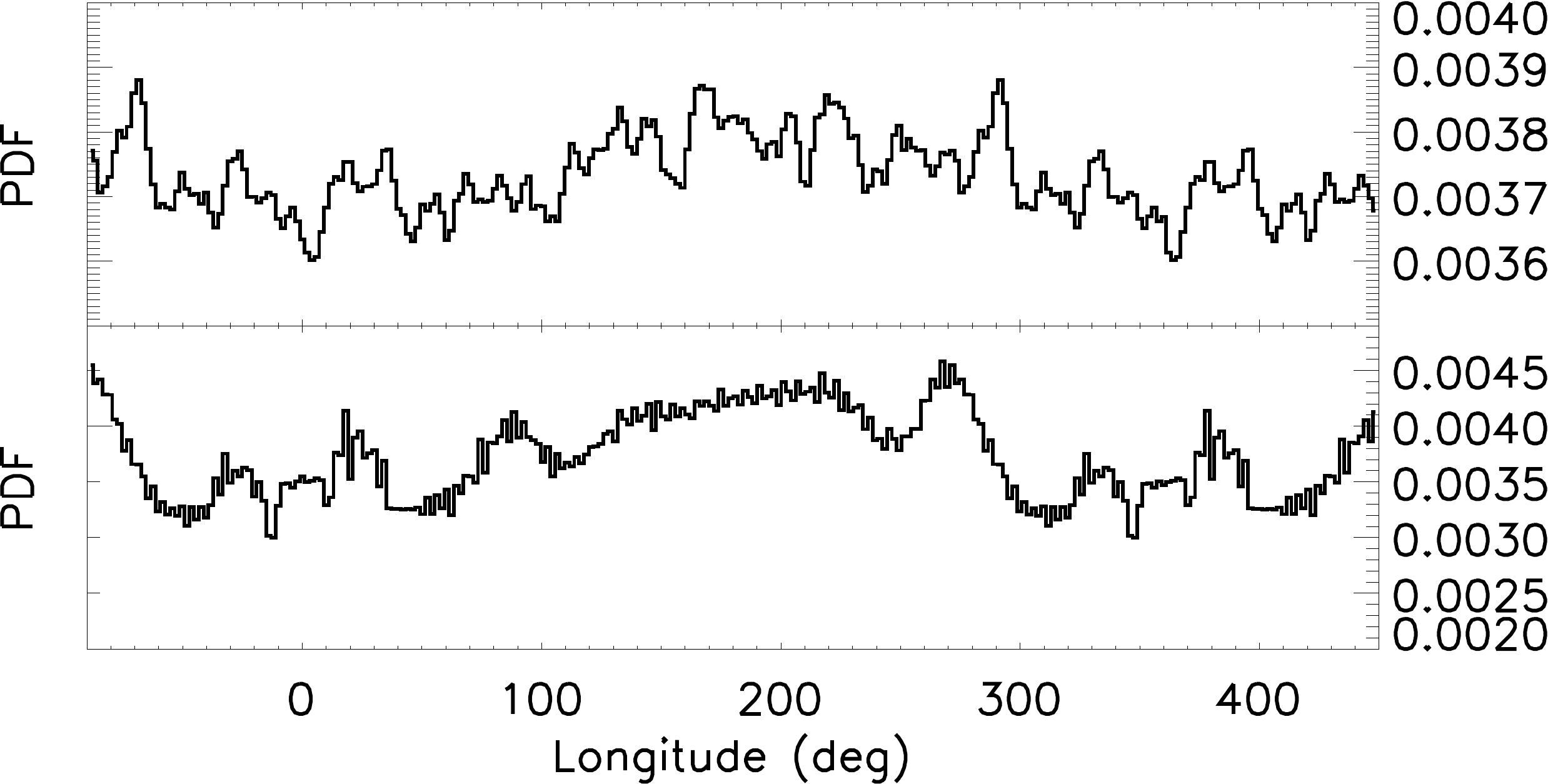}
\end{center}
\caption{Same as Fig.~\ref{rand}, for two migrating spot groups in longitude at a rate of $1^\circ$ per orbital cycle. The black dots in (a) and (b) are offset by +0.002 and +0.004 respectively, for visibility. {The black inclined lines show the trajectories of the geometric centres of input clusters.}} 
\label{mig_rand}
\end{figure}

Figure~\ref{mig_rand}a/b show the input and reconstructed maps of drifting clusters for $10^3$ simulations. The capability of {\tt DoTS} in tracking the spot clusters upon a randomly changing, homogeneous background of small-scale spots is evident. 
While the clusters are trespassing the primary eclipse zone, there is an apparent conglomeration in subsequent reconstructions, owing to insufficient sampling rate. The PDFs corresponding to the input and output maps are shown in Fig.~\ref{mig_rand}, which show similar large-scale distributions, mainly reflecting the time-averaged longitudinal coverage of the drifting spot clusters. 

To measure the drift rate of the spot clusters, we fitted lines to blindly selected sections of the tracks in Fig.~\ref{mig_rand}b (see also Sect.~\ref{sssec:drtime}) and estimated an average of 1.89 $\pm$ 0.16 deg/day, which is very close to the imposed rate, shown by continuous parallel lines in Fig.~\ref{mig_rand}b. Opposite longitudinal hemispheres exhibit modulation of mean spot occupancy in anti-phase, analogous to the `flip-flop cycles' observed for some active close binaries \citep{berdyugina05}. Here, we show that the effect can be \emph{solely} led by the steady azimuthal migration of long-lived spot clusters in the orbital frame of reference, owing to differential rotation.

Further tests with the same time range showed that up to three well-separated spot clusters can be recovered reliably, whereas four clusters are only partially recovered in most cases. Furthermore, rotationally invariant additional spots (e.g., polar caps) do not tend to yield spurious longitudes. 
We note, however, that mismatches in the stellar parameters (e.g., limb darkening coefficients) can yield spurious stationary patterns \citep[e.g.,][]{Jeffers2005}.

\subsection{Spot modelling: KIC 11560447} 
\label{spotmodelkic}

In long-term activity monitoring using photometry alone, the individual light curves should be normalised to the maximum brightness level, to reveal the absolute variations of the overall spot distribution. In our case, however, light curves for each orbital cycle should be normalised to their own maximum brightness levels, because the reconstruction algorithm of the {\tt DoTS} code automatically rescales the amplitudes of the synthetic light curves, without considering an absolute light level. Therefore, the spot filling factors obtained in this study measure merely the spot contrast distribution relative to the maximum brightness detected for each orbital revolution. On the one hand, the long-term variation of the global mean relative spot occupancy is meaningful as long as at least one orbital epoch with an immaculate visible hemisphere is present. On the other hand, the spot distributions in subsequent orbits are somewhat comparable with each other, because the emergence, evolution, and longitudinal migration of spot groups are traceable in consecutive reconstructions. Nevertheless, if a polar spot gradually develops over a number of orbital revolutions, its \emph{axisymmetric} (rotationally invariant) contribution would not be detected in our procedure, whereas sufficiently large equatorward spot extensions could be captured. 

In what follows, quantities plotted to display their azimuthal distributions are all functions of the longitude of the \emph{orbital frame}, where spot groups at latitudes not synchronised with the binary rotation period would shift in orbital phase at differing rates, as will be explained in Sect.~\ref{sssec:drtime}. 

\begin{figure}
\begin{tabular}{cc}
\includegraphics[width=3.2cm,angle=0]{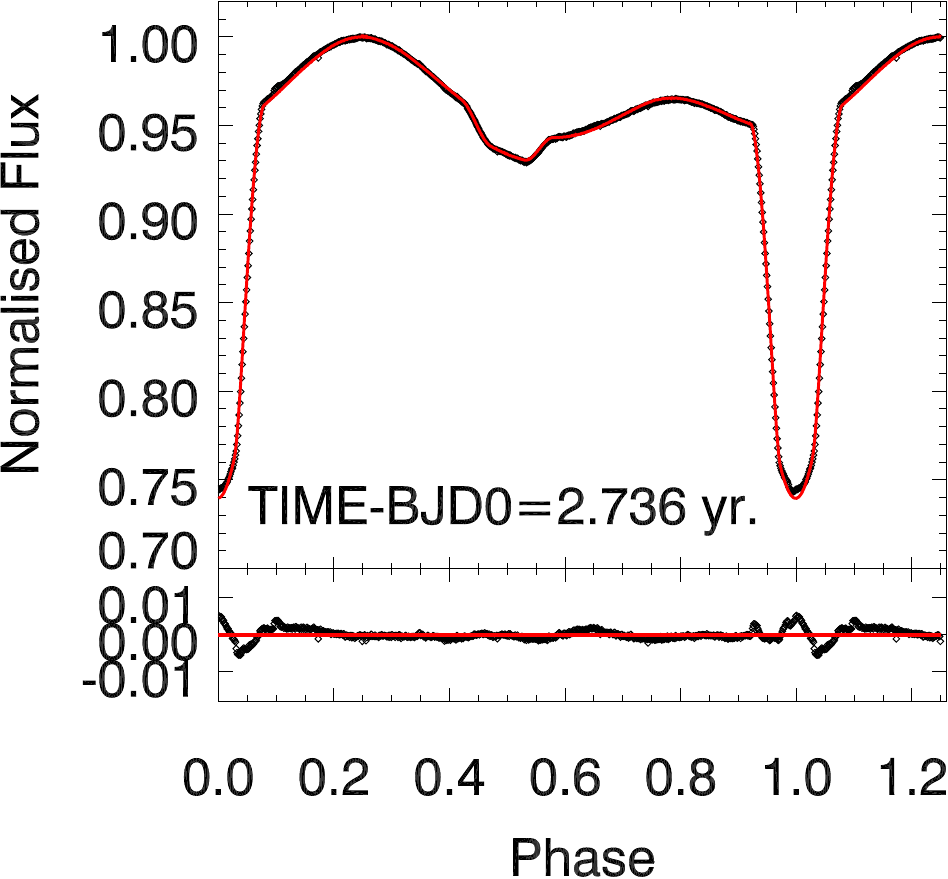}
\includegraphics[width=5cm,angle=0]{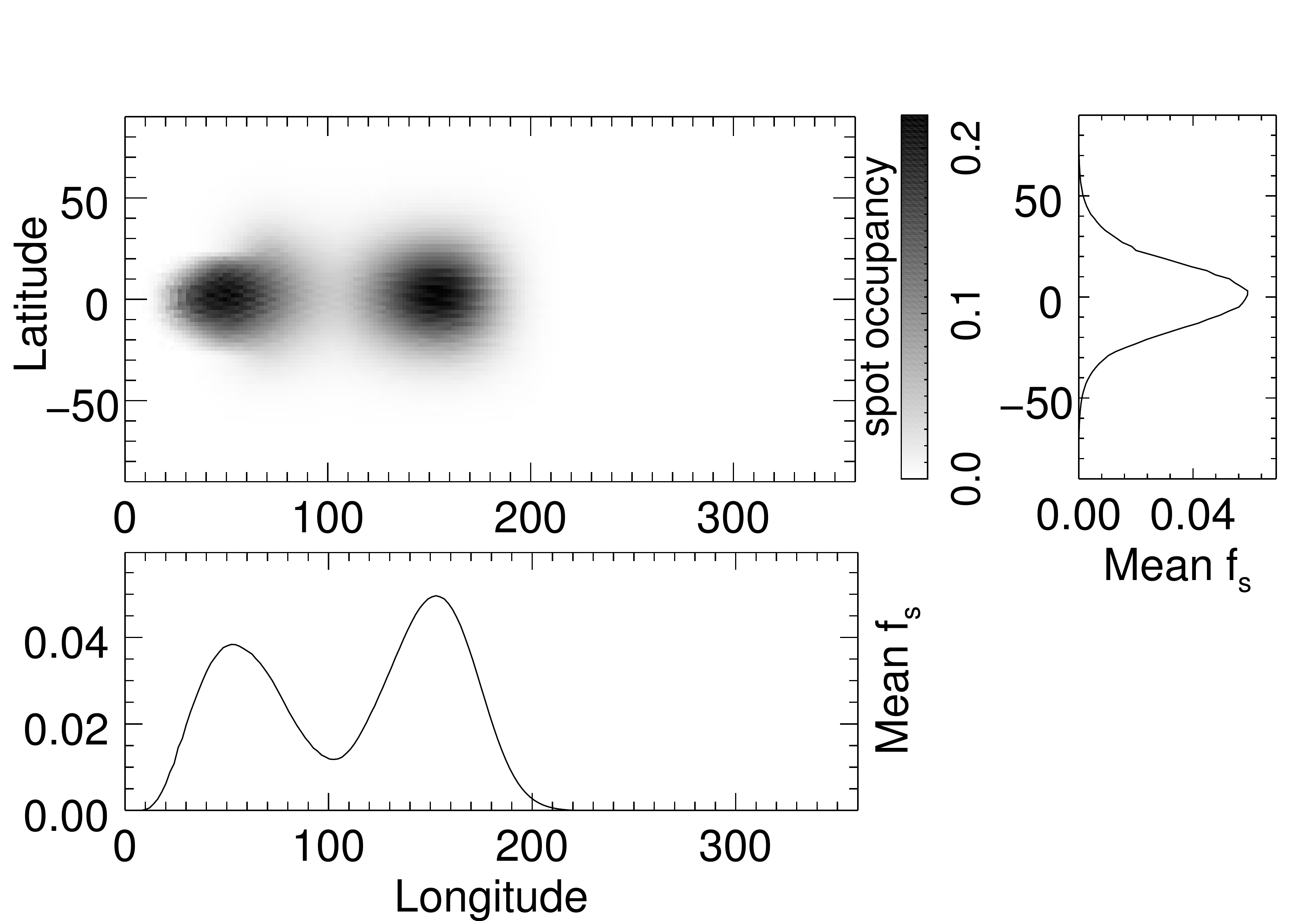}
\end{tabular}
\begin{tabular}{cc}
\includegraphics[width=3.2cm,angle=0]{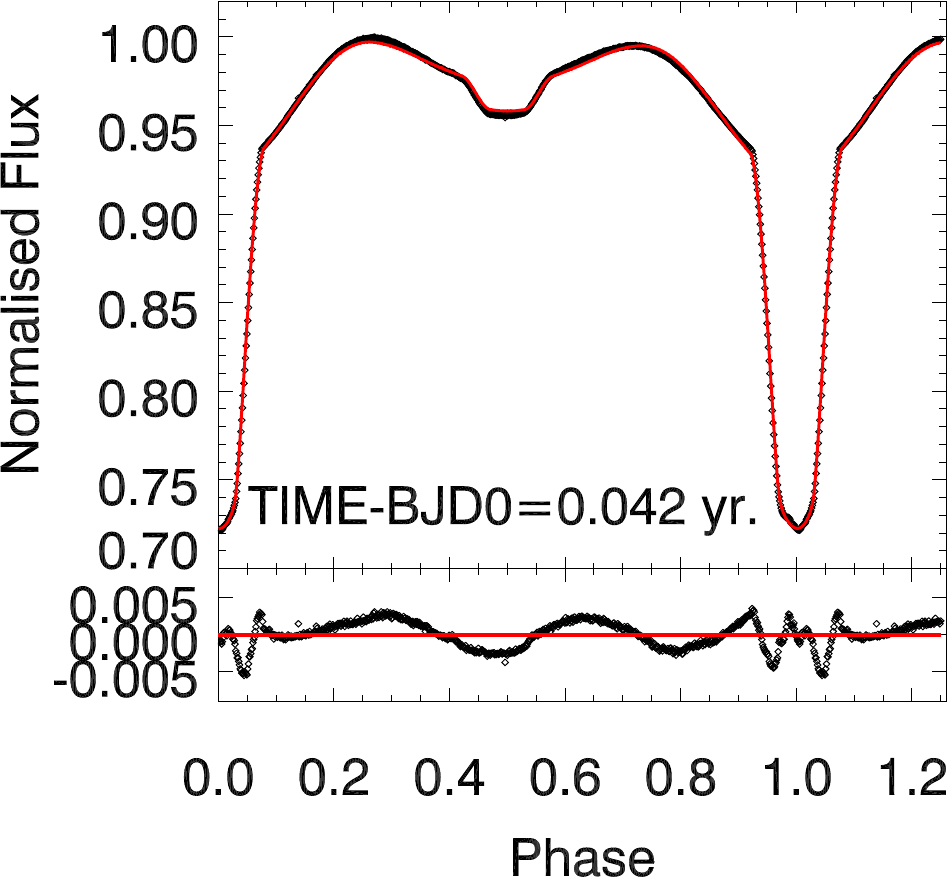}
\includegraphics[width=5cm,angle=0]{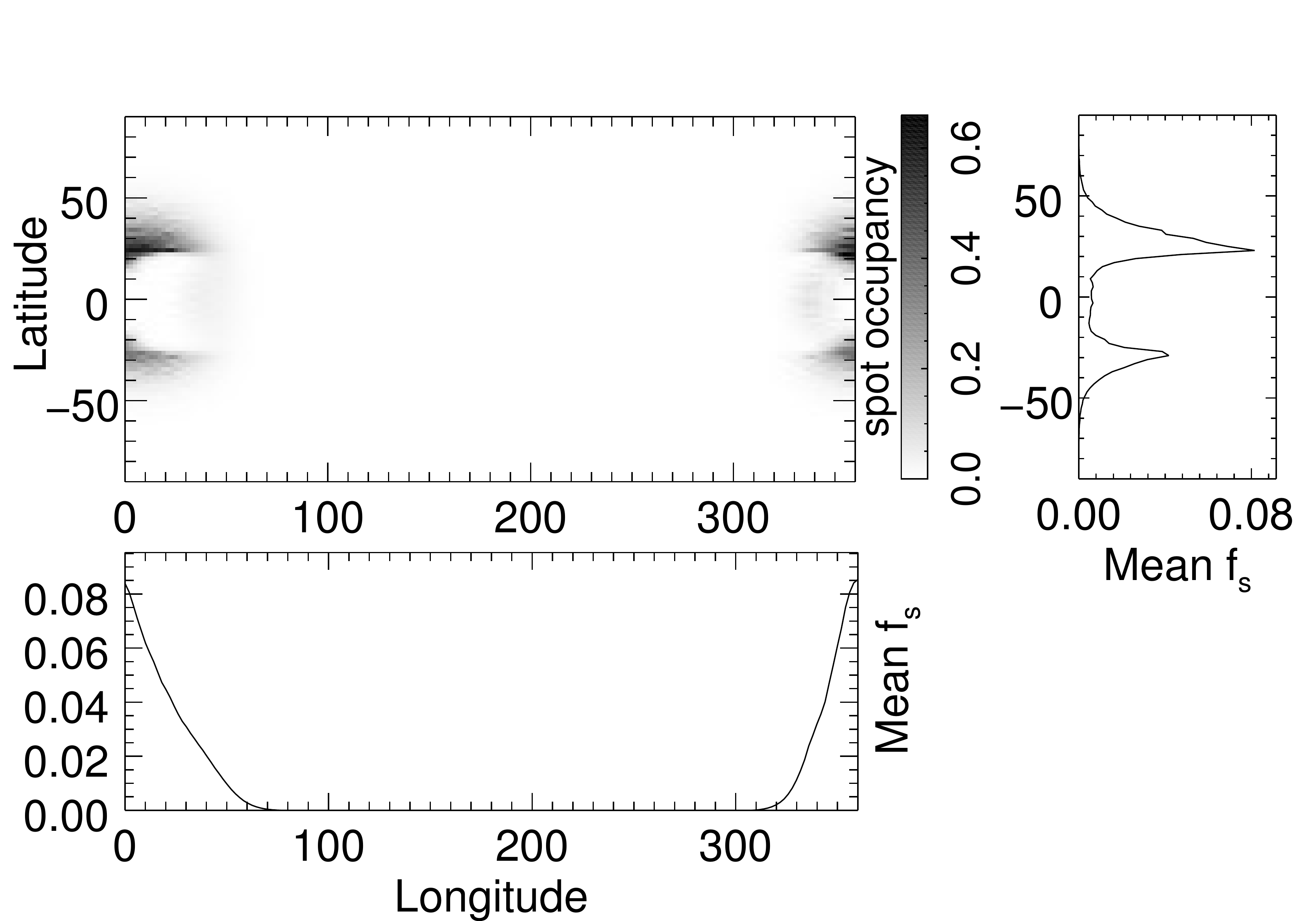}
\end{tabular}
\begin{tabular}{cc}
\includegraphics[width=3.2cm,angle=0]{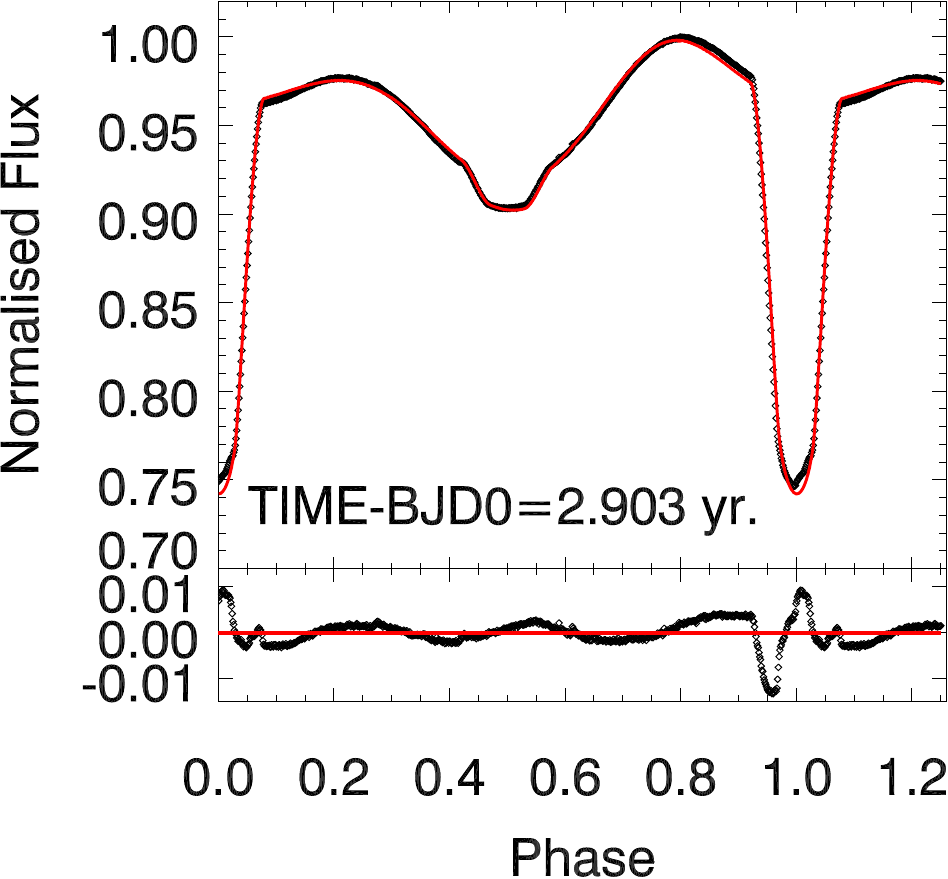}
\includegraphics[width=5cm,angle=0]{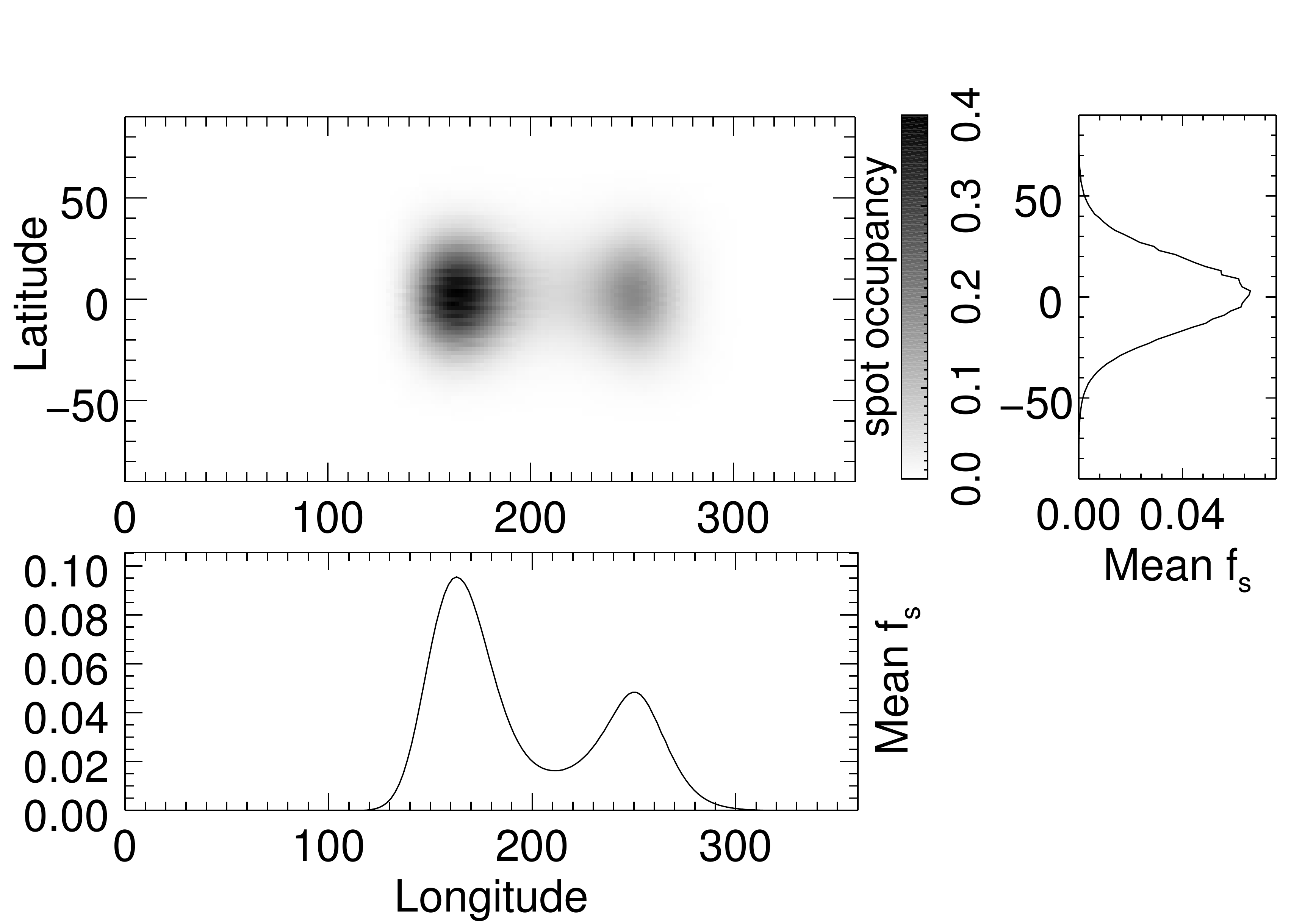}
\end{tabular}
\caption{Three light curves (left-hand panels, black dots for SC data, red line for the model) with the residuals from the model curve (at the bottom of each light curve) and the corresponding surface maps reconstructed (right-hand panels) at three different orbital epochs. Note that the middle panel represents the light curve modelling and the corresponding surface map for the 30$^{\rm th}$ cycle.}
\label{threemodel}
\end{figure}

Motivated by the numerical tests, we performed photometric mapping for the primary component of the system using the entire \emph{Kepler} data available for the system. We phased each light curve, using the light elements given in Eq.~\ref{ligel}, for both LC and SC datasets. Photometric limb darkening lookup tables required for the photometric mapping mode of {\tt DoTS} were generated by considering the \textit{Kepler} limb darkening coefficients by \citet{claret11}. We used {the} system parameters given in Table~\ref{table2} and a spot temperature of 4000~K as found from our {\tt PHOEBE} analysis. 
Three light curve models and the corresponding surface {reconstructions} on three different epochs are given in Fig.~\ref{threemodel}. The middle panel corresponds to the $30^{\rm th}$ cycle, for which the longitude of the reconstructed spot is well in accordance with the one found from the light curve modelling via the {\tt PHOEBE} code, which is based on forward modelling. We note that light curve inversions always place the spots near the equator for nearly edge-on systems, because the latitudinal information cannot be obtained \citep[see][and the references therein for a discussion of various inversion techniques]{berdyugina05}.

\section{Results}

\subsection{Activity in time and longitude}
\label{ssec:longdist}

Figure~\ref{fig:map} shows the longitudinal distribution of the {latitudinally averaged} relative spot filling factor, $f_s(\phi,t)$, as a function of time, {along with the} global and hemispheric means (right panel) and PDFs (lower panel) for LC and SC data. The resulting patterns of spot occupancy and hemispheric means for both datasets are in accordance for overlapping epochs.

A conspicuous feature seen in Fig.~\ref{fig:map} is the occasional emergence and decay of spotted regions in certain longitudinal intervals, such as the ones at median longitudes and time ranges of about ($200^\circ$, 0.7-0.9 yr), ($180^\circ$, 1.2-1.5 yr), ($150^\circ$, 2.2-2.5 yr), ($225^\circ$, 3.1-3.5 yr). 
Almost all the regions of relatively high spot {filling factor drift} towards increasing longitudes, which will be investigated in Sect.~\ref{sssec:drtime}.

The mean spot filling factor in opposite longitudinal hemispheres centred at $\phi_0=270^\circ$ and 90\textdegree ~vary quasi-periodically with a phase difference of 180\textdegree. A similar pattern was mentioned by \citet{ioan2016}, for the single K-dwarf \emph{Kepler}-210, which has a rotation period of about 12 days. Our simulation of drifting spot clusters {indicates} that such alternating variations in hemispheric spot coverage can be simply driven by differential rotation of the dominant spot regions. However, intrinsic variations led by spot emergence and decay can also play a role. This possibility deserves further studies involving numerical simulations.

Longitudinal PDFs based on LC and SC data are shown in Fig.~\ref{fig:map}. In spite of the smearing effect of varying rotation rates seen in {the time-longitude diagram}, the PDFs {indicate} two `preferred' longitude zones, centred roughly around $0^\circ$ and $140^\circ$. 
No such clear preference at those longitudes was seen in the simulation of two drifting clusters (Fig.~\ref{mig_rand}c, lower panel), which has a drift rate that is comparable with those of several drifting patches in Fig.~\ref{fig:map}.

As a model-free test for both the presence of preferred longitudes and the performance of {\tt DoTS}, {we calculated the deviation of each light curve from an immaculate model, by the following procedure: We first converted the count rates for all light curves to \emph{Kepler} magnitudes, then to arbitrary fluxes. Taking the model light curve of the $30^{\rm th}$ orbital cycle (Fig.~\ref{lcfit}), we eliminated the effect of the spot which was fit using {\tt PHOEBE}. To serve as a reference for the observed light curves, we rescaled the resulting normalised, `spotless' model light curve such that it partially matches with the light curve $F_{30}(\phi)$ (the scaling was made only to fit the part around the secondary minimum, because the 30th cycle has a spot near primary minimum). Calling this rescaled model curve $F^{\rm M}_{30}(\phi)$, we calculate} the \emph{flux contrast}, 
\begin{equation}
C_F(\phi)\equiv\frac{F^{\rm M}_{30}(\phi)-F_i(\phi)}{F^{\rm M}_{30}(\phi)}, 
\end{equation}
where $F_i(\phi)$ is the light curve (in flux units) of the \emph{i}th orbital cycle, {so that positive contrast means spot darkening. The corresponding magnitude difference is $m_{30}-m_i = 2.5\log(1-C_F)$.}
Free from any artefacts due to the reconstruction procedure, $C_F$ is a differential measure for the combination of rotational modulation of radiative flux owing to surface inhomogeneities, and secular variations in system brightness, all relative to our reference model of cycle 30. 

{The resulting $C_F(\phi,t)$ is shown in Fig.~\ref{lcocvar}, along with hemispheric and full-sphere rotationally averaged values, and the longitudinal PDF.}
The overall spatio-temporal match between the flux contrast and the reconstructions in Fig.~\ref{fig:map} validates the sensitivity of {\tt DoTS} solutions especially to the {large-scale} spot activity {leading to rotational modulation}. 
The $\phi$-independent (axisymmetric) variations are in line with the variation in {rotationally averaged system} brightness (Fig.~\ref{maxs_fs}a){, \emph{i.e.}, the secular increase in the total flux contrast is visible, especially during the final year}. 
The PDF of flux contrast {supports} the presence of the two {mildly} preferred longitudes centred at $\phi_0\sim 0^\circ$ and $\phi_0\sim 140^\circ$ (same as for reconstructions). The asymmetric {shape of each peak} is consistent with the possibility that flux emergence occurs preferentially at these latitudes, but then the prograde drifts, regardless of their origin, flatten the right wings of both PDF peaks. {However, one should also consider that the primary eclipse is sampled by only 3-4 data points in each long-cadence light curve, so the PDF peak at these phases is possibly affected by the lack of data at the middle of the eclipse. 
The hemispheric mean contrast levels show essentially the same behaviour as do the corresponding $\langle f_s\rangle$: the dominant hemisphere is interchanged, as the spotted regions emerge with a preferential longitudinal pattern} and drift prograde in the orbital frame. 
In the final months, the amplitude of the hemispheric modulation is still visible, but highly suppressed (Fig.~\ref{lcocvar}, right panel).

\subsection{Hemispheric and global variations}
\label{ssec:cycles}

\begin{figure}
\hskip-3.mm
\includegraphics[width=20pc]{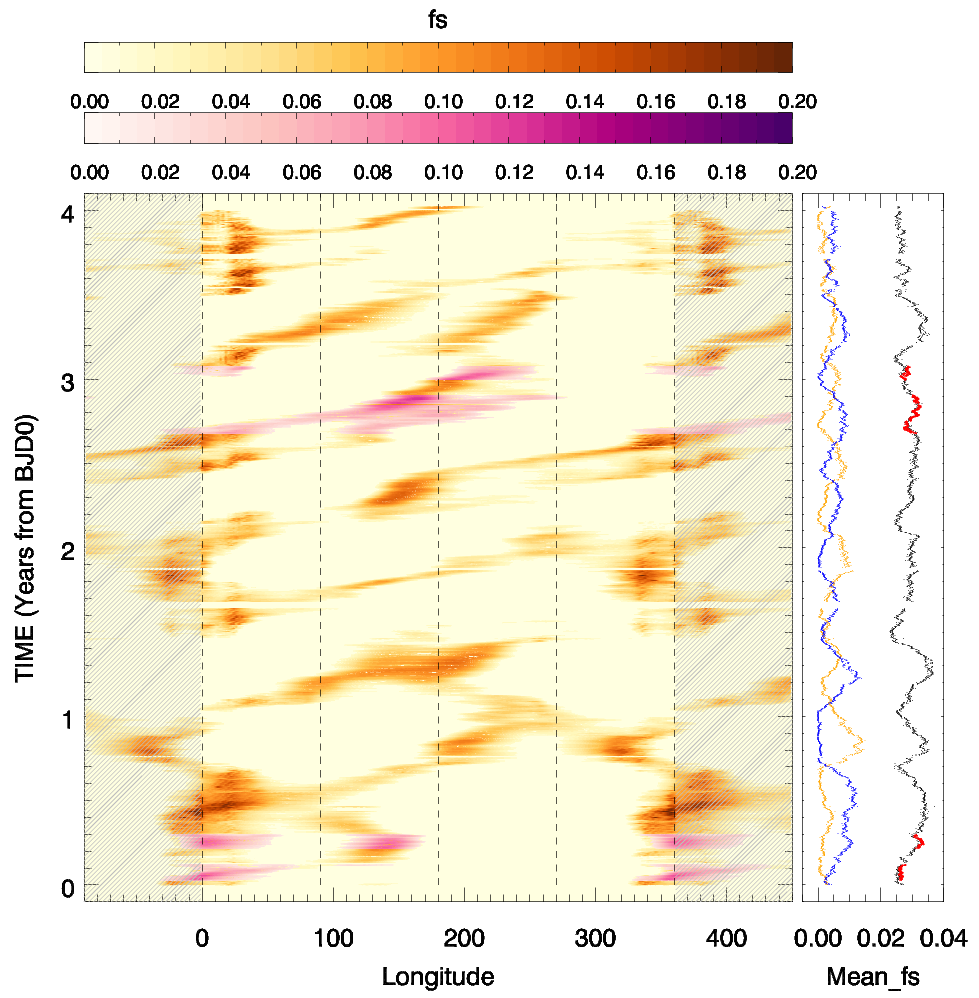}\\
\hskip+3.mm
\includegraphics[width=17pc]{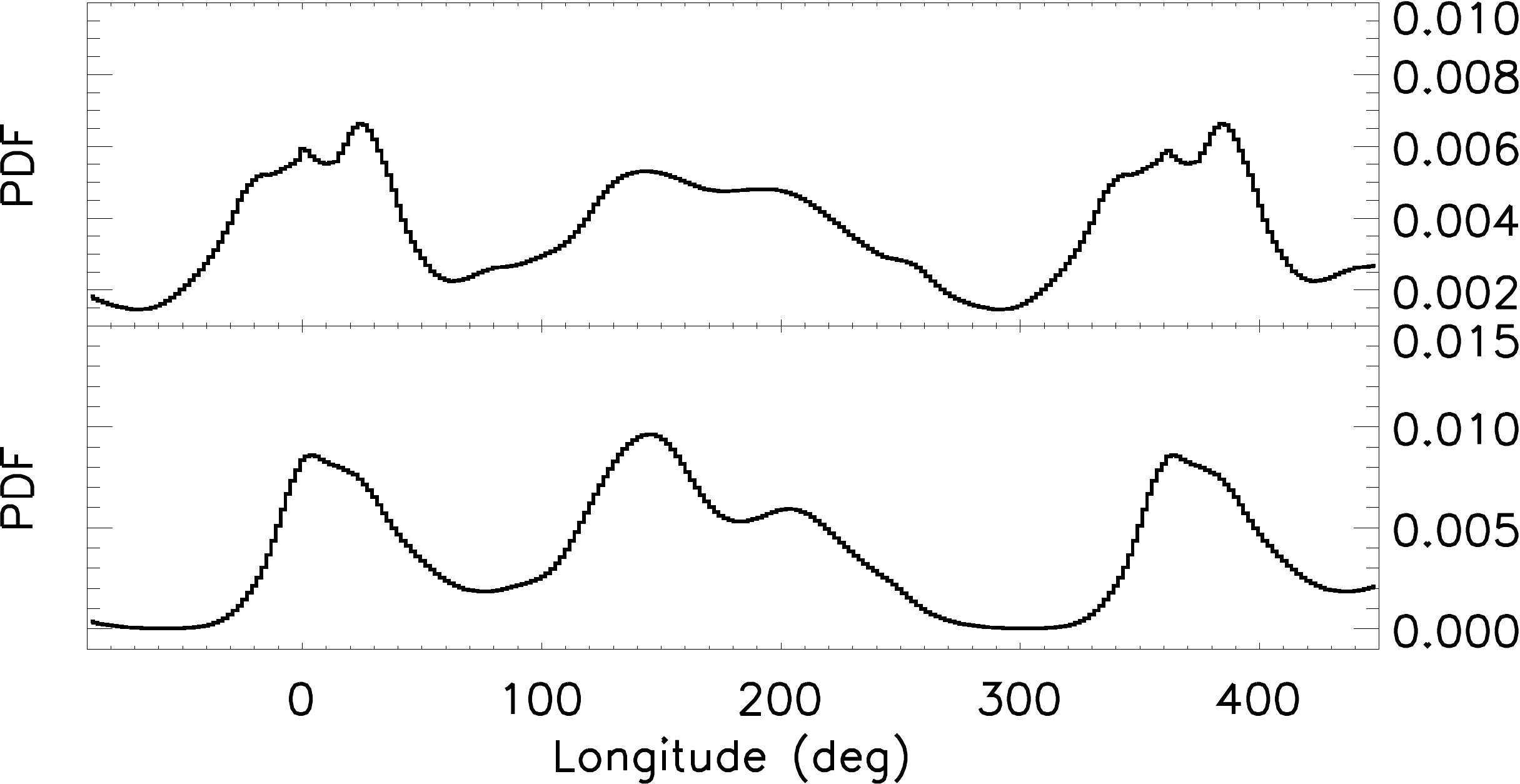}
\caption{{\emph{Upper panel:}} Time-longitude diagram of the relative spot filling factor $f_s$ of the primary component of KIC 11560447, covering $\sim 2800$ orbital cycles. The upper and lower colour bars represent the LC and SC data, respectively. \emph{Right panel}: the global mean $\langle f_s\rangle_\phi$ (offset by +0.02) for LC (black) and SC (red) data; the hemispheric mean $f_s$ (offset by +0.005) for central longitudes 270\textdegree ~(orange) and 90\textdegree ~(blue). {\emph{Lower panel:} PDFs for the LC (upper plot) and SC data (lower plot).} }
\label{fig:map}
\end{figure}

\begin{figure}
\begin{center}
\includegraphics[width=20pc]{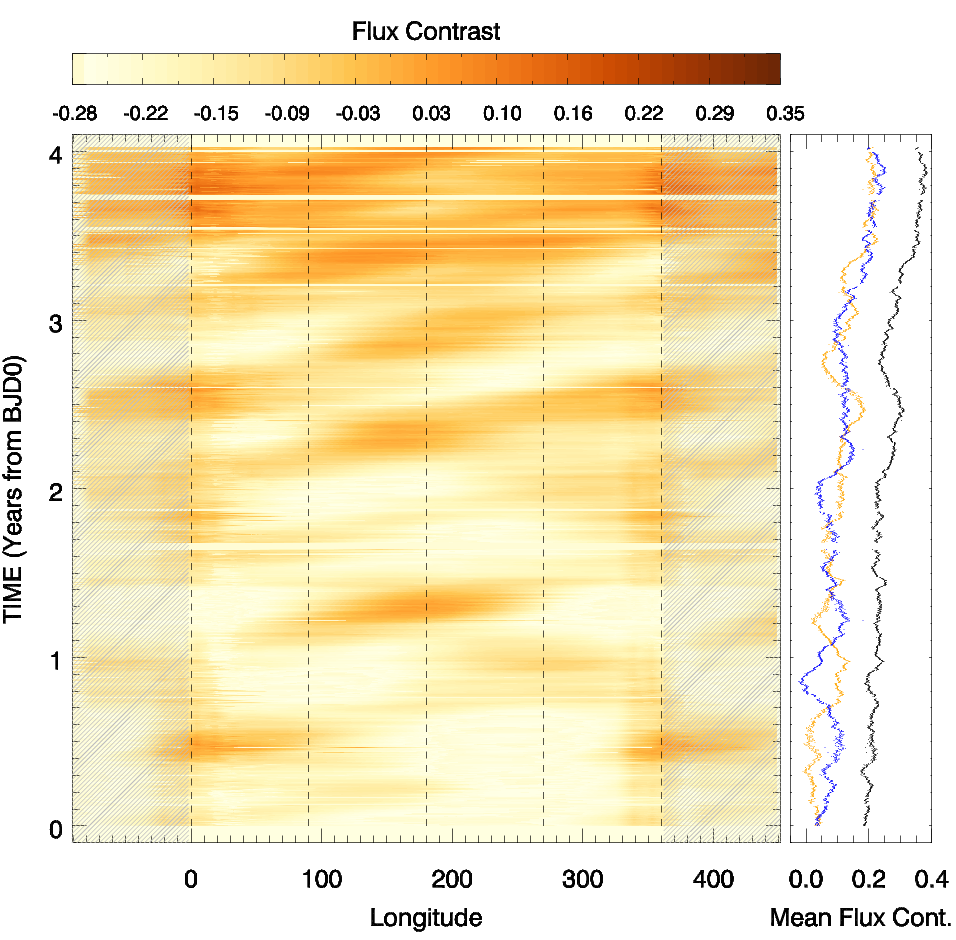} \\
\hskip-16.0mm \includegraphics[width=17pc]{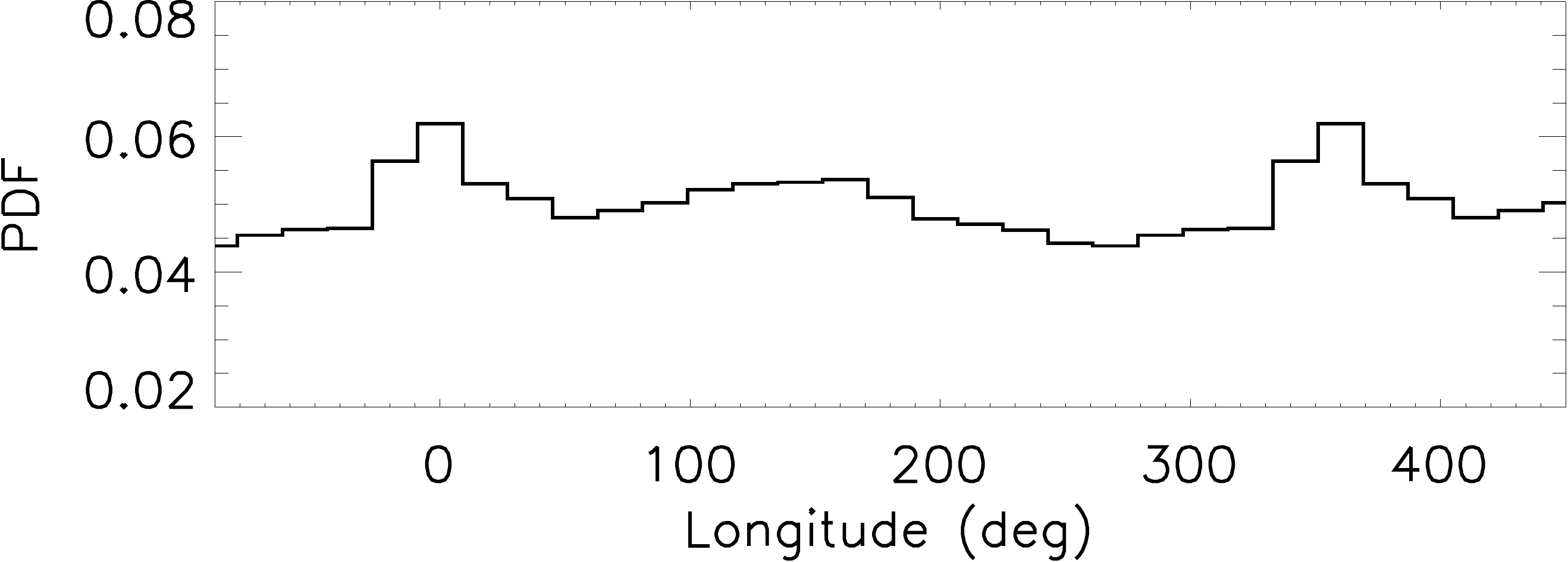}
\caption{Time-longitude diagram for the radiative flux contrast with respect to the {model} light curve of $30^{\rm th}$ orbital cycle, {from which the spot modulation was eliminated}. Both axisymmetric and nonaxisymmetric components emanating from surface inhomogeneities are visible in this figure. Similar to that of Fig. ~\ref{fig:map}, the right panel shows the global (black) and hemispheric averages (orange and blue) of $C_F$. The lower panel shows the PDF of the entire data.}
\label{lcocvar}
\end{center}
\end{figure}

\begin{figure*}
\begin{center}
\includegraphics[width=\linewidth]{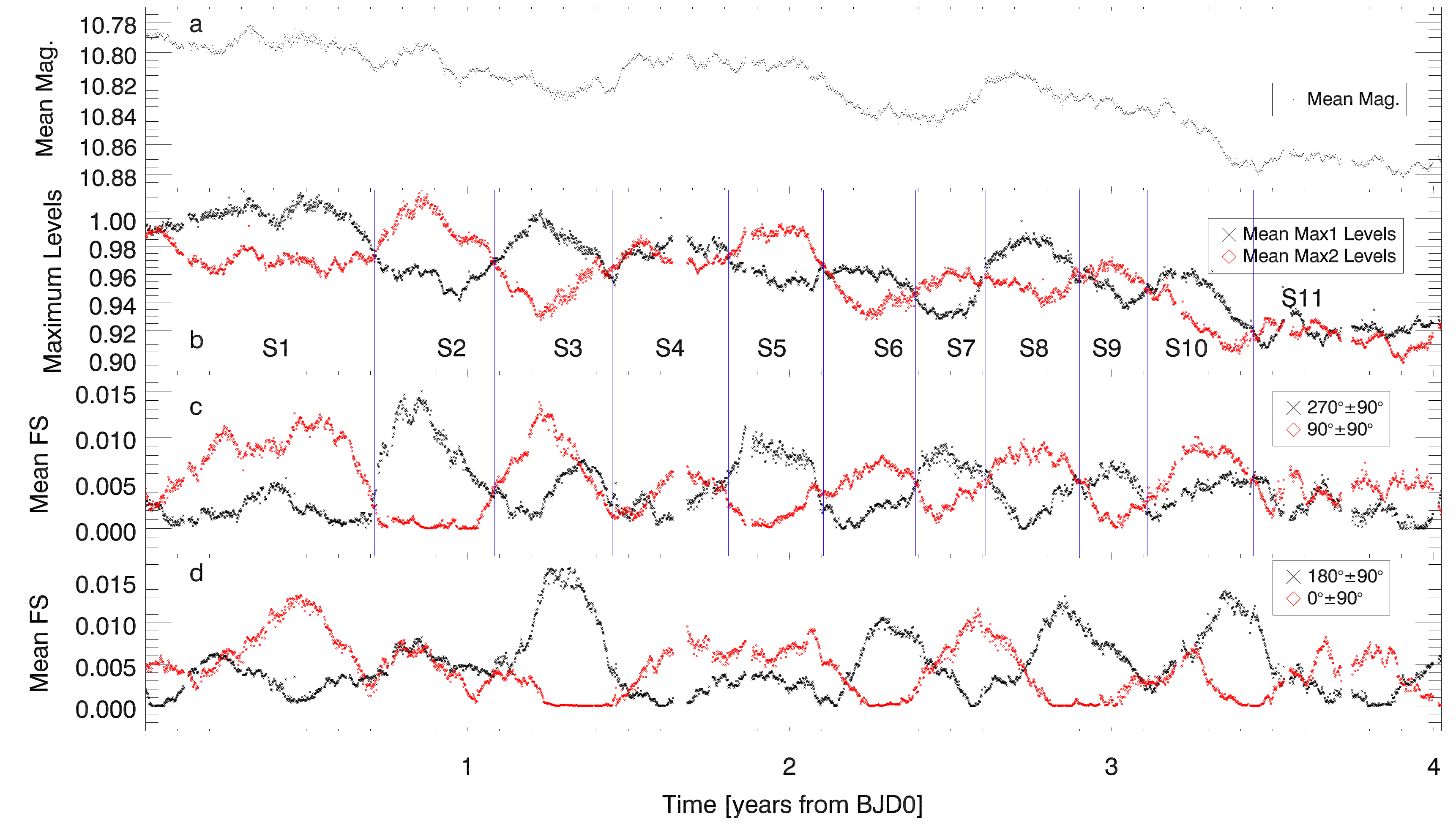}
\caption{(a) {Orbital averages of} \textit{Kepler} magnitudes, (b) the primary (black) and secondary (red) maximum levels. Panels (c) and (d) show the relative spot filling fractions {of the reconstructions} in the two opposite hemispheres 
(centred at 270\textdegree (black) vs. 90\textdegree (red) for panel (c); 180\textdegree (black) vs. 0\textdegree (red) for panel (d), respectively).} 
\label{maxs_fs}
\end{center}
\end{figure*}

We compare in Fig.~\ref{maxs_fs} {hemispheric brightness variations with the corresponding spot filling factors.
As seen in terms of the flux contrast above, the rotationally averaged} brightness (Fig.~\ref{maxs_fs}a) shows a secular dimming, with occasional temporary depressions lasting for a few months. 
We define the levels of the {first and second} maxima, as the normalised fluxes averaged over orbital phase intervals [0.10,0.45] and [0.55,0.90], respectively, corresponding to longitudinal intervals [198\textdegree,324\textdegree] and [36\textdegree,162\textdegree]. The variation of the {first and second maximum} levels are shown in Fig.~\ref{maxs_fs}b, and the relative mean spot filling factor of two sets of opposite longitudinal hemispheres (with centres $\phi_0$ at 270\textdegree ~vs. 90\textdegree and 0\textdegree ~vs. 180\textdegree) are shown in Fig.~\ref{maxs_fs}c and Figure~\ref{maxs_fs}d, respectively. The correspondence between the humps in the maximum levels and those in the spot occupancy of the relevant hemispheres is directly visible from the plots {(Cf. Fig.~\ref{fig:map})}. These humps occur more or less in parallel with the depressions in the {rotationally averaged} brightness (eg., for S2, S3, S6, S7, S11), {which indicates hemispheric preference in spot emergence, in agreement with the histograms in Figs.~\ref{fig:map}-\ref{lcocvar}.}
We have found a weak anticorrelation between the {first} maximum level and the mean spot coverage of the hemisphere $\phi_0=270^\circ$ (correlation coefficient $-0.52$). There is a stronger anticorrelation (coefficient $-0.73$) between the {second} maximum level and the mean spot coverage of the {corresponding} hemisphere $\phi_0=90^\circ$. 
Considering the entire dataset, we did not find a significant correlation between the total brightness and the relative spot coverage. One reason is that the latter quantity does not correspond to the absolute spot occupancy, but reflects the spot occupancy relative to the maximum of each light curve, which might well be contaminated by axisymmetric spot distributions such as a polar cap, as discussed in Sect.~\ref{ssec:longdist}.


\begin{figure}
\begin{center}
    \includegraphics[width=\columnwidth]{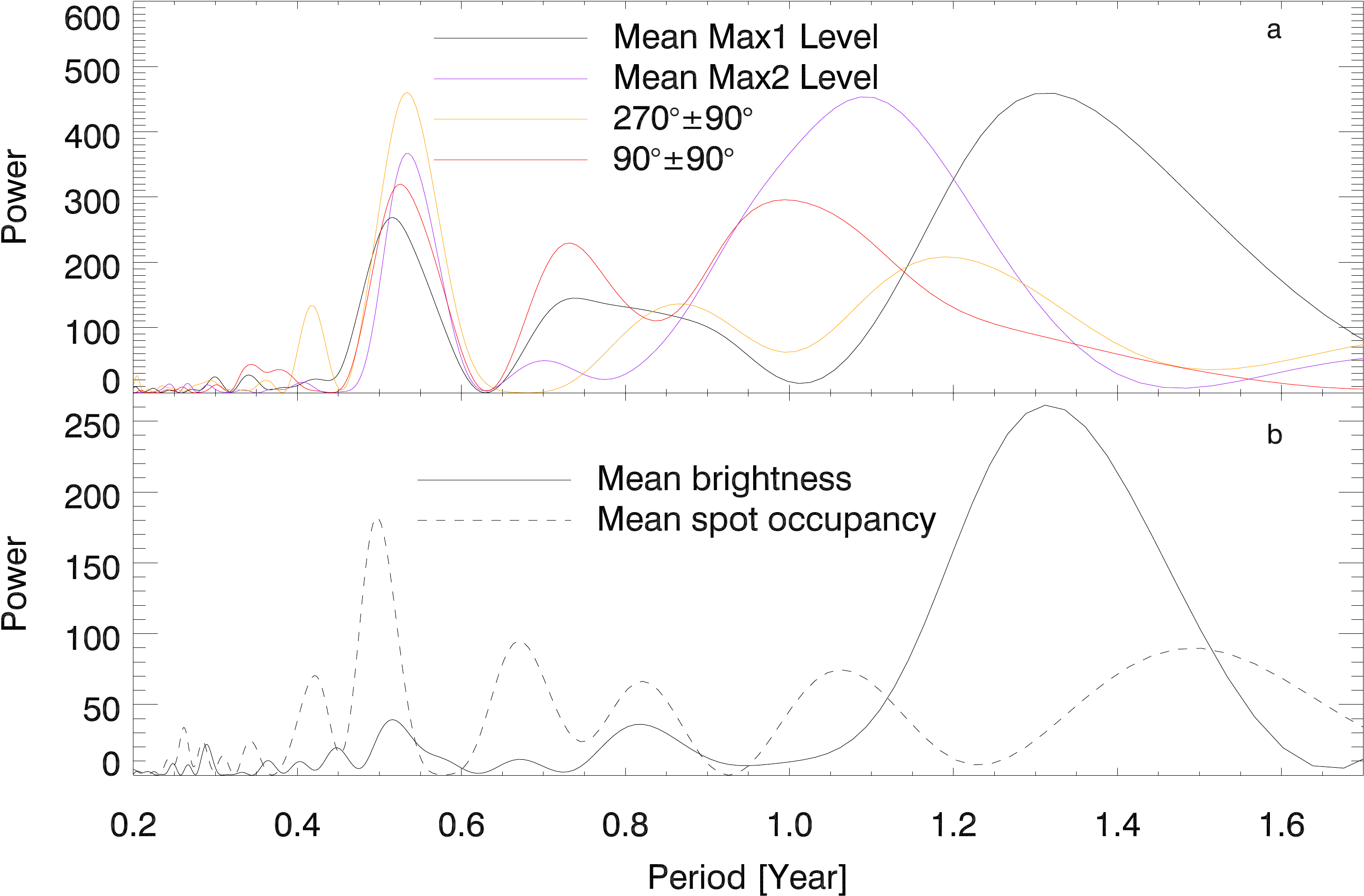}
    \caption{Lomb-Scargle periodograms for the primary and secondary maximum levels, spot occupancy of the hemispheres centred at longitudes 270\textdegree and 90\textdegree (upper panel); LS periodograms for the brightness and the spot coverage (lower panel).}
\label{fig:periodograms}
\end{center}
\end{figure}

To assess the link between the variations of the maximum levels and the spot occupancy of the corresponding hemispheres, we show the corresponding Lomb-Scargle periodograms in Fig.~\ref{fig:periodograms}a and the {rotationally averaged} brightness and spot occupancy in Fig.~\ref{fig:periodograms}b. All the periodograms {peak prominently} at about 0.5 year, {which we attribute to} recurrent emergence and decay of large-scale spot regions with typical lifetimes of half a year, {as can be seen in Fig.~\ref{fig:map}}. 
In the range 0.7 to 1.4 years, there are a number of peaks that show {partial} correspondence among the maximum levels, spot coverage, and the total brightness. 
These periods can be related to the combined effect of recurrent emergence and the relative rotation of spotted regions in the orbital frame. 
Another possible effect related to differential rotation among active latitudes (2-frequency beats) will be discussed in Sect.~\ref{sssec:drfreq}. 

\subsection{Surface differential rotation}
\label{ssec:dr}

\subsubsection{Time domain: phase shifts in spot patterns}
\label{sssec:drtime}

The streaking patterns of spot concentrations in Fig.~\ref{fig:map} have {the interesting property that} they drift preferentially towards increasing longitudes, i.e., in a \emph{prograde} sense, with very few {features} moving in the retrograde sense. 
In some cases, spot regions with different longitudinal {drift} rates seem to overlap.
{Our longitudinal grid is at rest in the} \emph{orbital frame}, 
where spot groups not synchronised with the binary rotation period would shift in orbital phase. 
Possible effects that can {manifest themselves as such streaking patterns} are ($i$) surface differential rotation, ($ii$) {successive} emergence of multiple magnetic flux tubes {( out of the same toroidal flux bundle)} in the prograde direction, and ($iii$) {prograde} azimuthal propagation of dynamo waves \citep{cole14}. 
Following a simple and natural approach, we interpret the drifts with various rates around a few degrees per day as the imprint of latitudinal differential rotation, keeping in mind that systematic effects driven by magnetic fields dynamics and radial shear can co-exist, though their consideration is obviously beyond the scope of the current study. 

Our assumption of latitudinal differential rotation implies that the {spots emerging and decaying} at a fixed longitude {would be} at rest in the orbital frame, {which is synchronised with the latitude of such spots, called} $|\lambda_{\rm sync}|$. With increasing latitudinal distance from $|\lambda_{\rm sync}|$, the patterns of spot evolution would be increasingly inclined towards prograde (increasing) or retrograde (decreasing) orbital longitudes, depending on the {location of $|\lambda_{\rm sync}|$, the sign of $\alpha$ in Eq.~\ref{eq:sdr}, and the latitudinal distribution of spots}. 

In Table~\ref{tab:sync} we categorise the sense of longitudinal drift for spot groups under the effect of {either} solar-like {($\alpha>0$) or} anti-solar ($\alpha<0$) {SDR} patterns, for two cases for which $|\lambda_{\rm sync}|$ is close to the equator {(low)} and to the pole {(high)}. Assuming that spot groups {(at latitudes $\lambda$) are distributed} more or less evenly on both sides of $|\lambda_{\rm sync}|$, the drift direction for the majority {(minority)} of spot groups is written in \emph{italic} (normal) characters. 
In our case of mostly prograde drifts (Fig.~\ref{fig:map}), the scheme would imply the following two possibilities: $(i)$ solar-like surface shear with a {high-latitude $|\lambda_{\rm sync}|$}, $(ii)$ anti-solar shear with a low-latitude $|\lambda_{\rm sync}|$. In the first case, streaking patterns with lower slopes are located at high latitudes, and the most rapidly shifting spots are at the lowest active latitudes. 
When there is a significantly uneven latitudinal distribution of spots, the distribution of prograde and retrograde spots could differ from the basic scheme presented here. 

\begin{table}
\caption{Sense of longitudinal drift under different assumptions.}
\label{tab:sync}
\centering
\begin{tabular}{llll}
\hline
SDR 	&	$|\lambda_{\rm sync}|$	&	\multicolumn{2}{|c|}{Drift direction}\\
\hline
 &	& $|\lambda|$ < $|\lambda_{\rm sync}|$	&	$|\lambda|$ > $|\lambda_{\rm sync}|$ \\
 \hline
solar & low & {\it prograde} & {retrograde}\\
solar & high & {prograde} & {\it retrograde}\\
anti-solar & low & {\it retrograde} & {prograde}\\
anti-solar & high & {retrograde} & {\it prograde}\\

\hline
\end{tabular}
\end{table}
\begin{figure}
\begin{center}
\includegraphics[width=\columnwidth]{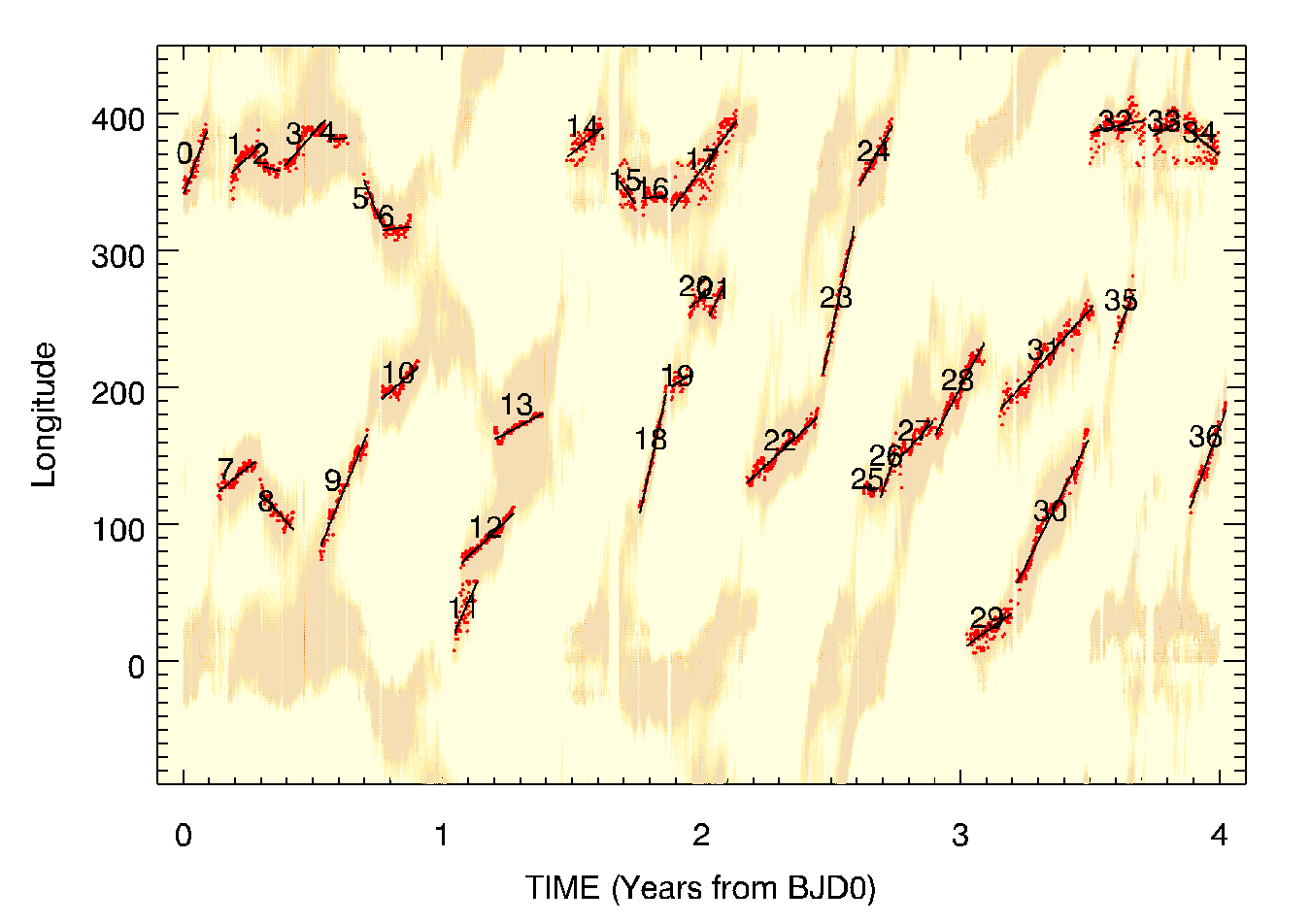}
\caption{Measurements of longitudinal drift rates using the LC spot filling fractions (Cf.~Fig.~\ref{fig:map}). The shaded regions correspond to spot filling factors higher than 0.025, the red dots to the median longitudes through which the black lines are fit.} 
\label{long_spots}
\end{center}
\end{figure}
To measure the rates of longitudinal drifts, we filtered the spot occupancy map (Fig.~\ref{fig:map}) with a lower cutoff {filling factor of} 0.025. 
Figure~\ref{long_spots} shows {37} linear fits {using LC data, and} 12 fits for the SC data, by manually choosing start/end times and longitudes out of the filtered data (shaded regions). For each time step, {the} procedure finds the median points (red dots) and fits a line to these points (black-coloured), {using linear least-squares method}. 

Table~\ref{table3} shows the results of the procedure, namely the median time (yr), lifetime (days), median longitude and longitudinal extent (degs), relative rotation rate {with respect to the orbital rate, $\Delta\Omega_i=\Omega_i-\Omega_{\rm orb}$} and the standard deviation in longitude of the median points from the linear fit. The detected range of relative rotation rate is {$[-1.35,2.44]$, which spans $3.79^\circ$/d} between the lowest and highest rates detected. We interpret this value as a rough estimate for the lower limit of surface differential rotation. In terms of relative differential rotation, $\Delta\Omega_i/\Omega_{\rm orb}$, this lower limit amounts to about 0.005. 

\begin{figure}
\centering
\includegraphics[width=\columnwidth]{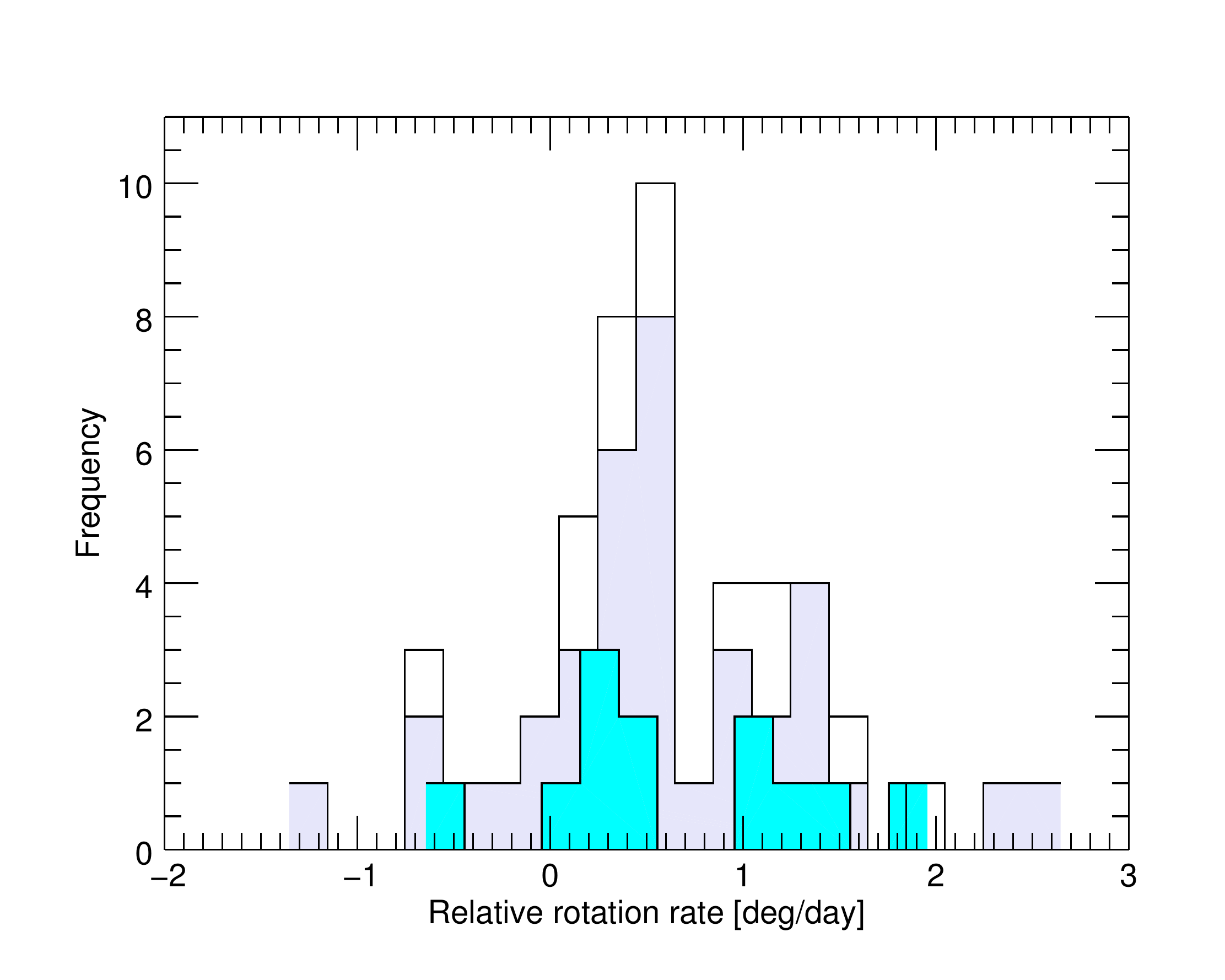}
\caption{Histograms of rotation rate in the orbital frame, based on the LC (light-shaded) and SC (dark-shaded {turquoise}) data. The SC histogram lies completely within the LC one. The sums of both data sets are also shown with a full line (unshaded).}
\label{fig:histomed}
\end{figure}

Figure~\ref{fig:histomed} shows histograms for $\Delta\Omega_i$, based on LC ($N_{\rm LC}=37$ measurements) and SC ($N_{\rm SC}=12$) data separately, and their sum. There is an overlap of a few regions in the two datasets (see Table~\ref{table3}). From the LC dataset, we find a mean relative rotation rate of {$\bar{v}\simeq 0.555^\circ$/d} with the standard deviation of the mean as $\sigma_{\bar{v}}/\sqrt{N_{\rm LC}}\simeq 0.127^\circ$/d. Based on the {SDR} hypothesis, the histograms show that the majority of spot regions {that we considered} are not far from $\lambda_{\rm sync}$, and/or that $\Delta\Omega/\Omega$ is significantly lower than on the Sun. 
We have also carried out an independent fit procedure using all the points exceeding the threshold (instead of median points) and obtained very similar {histograms with those in Fig.~\ref{fig:histomed}}. 

\subsubsection{Frequency domain: the latitudinal rotation pattern}
\label{sssec:drfreq}
\begin{figure}

\begin{center}
 \includegraphics[width=\columnwidth]{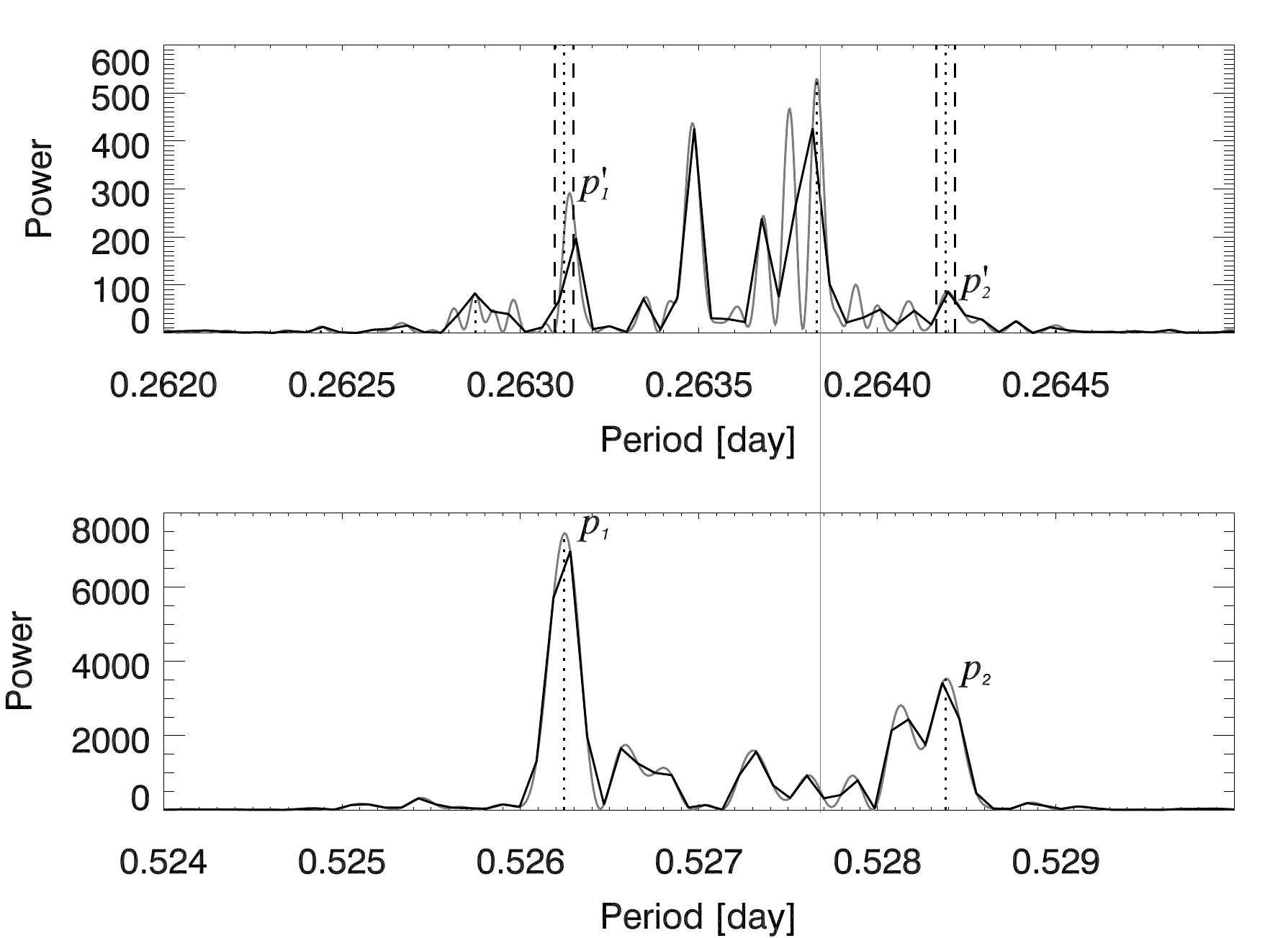} \\
\end{center}
\vspace{-6.5cm} \flushleft (a) \vspace{5.5cm} 
\begin{center}
\includegraphics[width=\columnwidth]{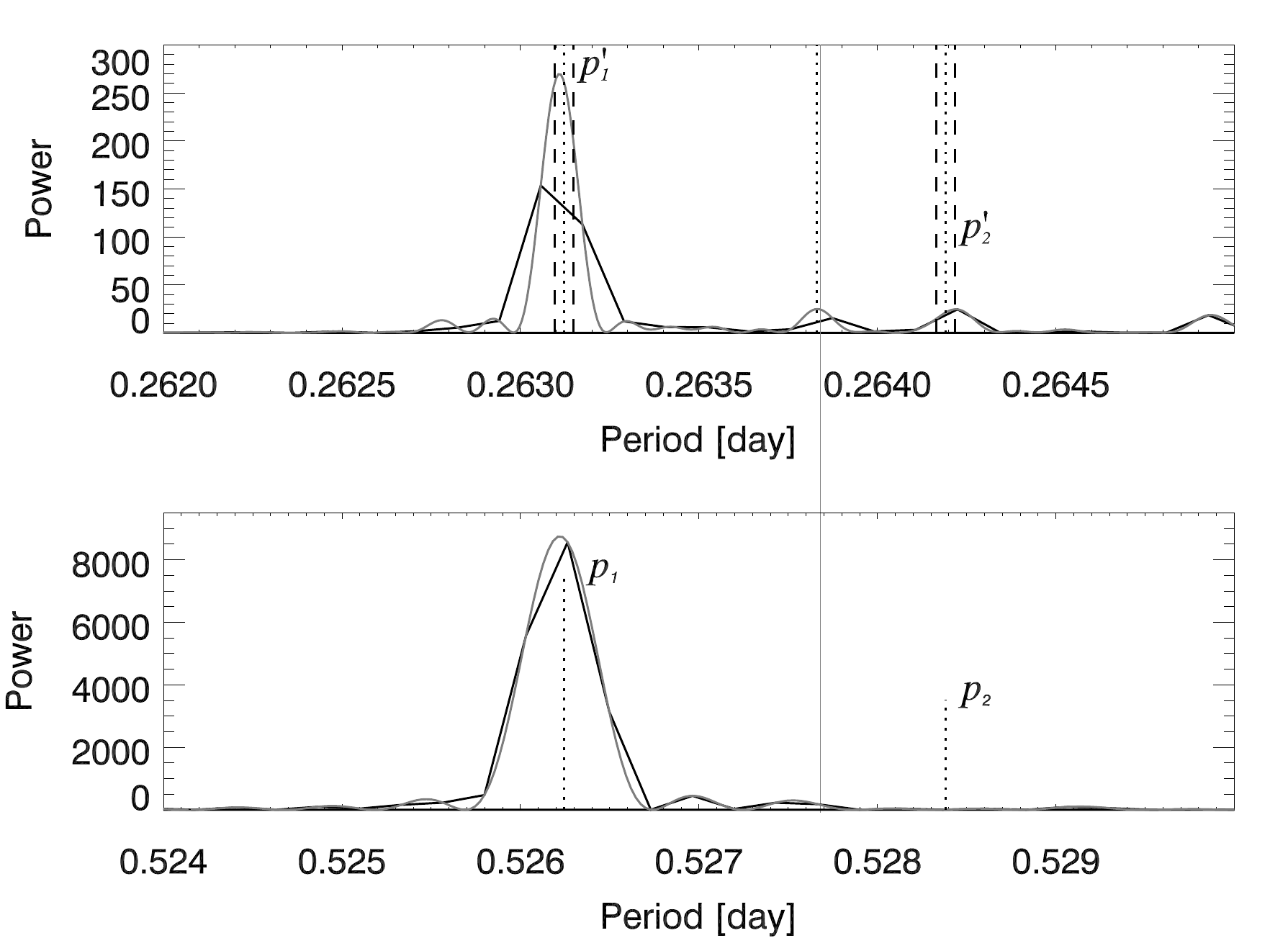}
\end{center}
\vspace{-6.5cm} \flushleft (b) \vspace{5.5cm} 
    \caption{(a) Lomb-Scargle periodogram for {model-subtracted LC data; grey curves are for an oversampling factor of 10}. The orbital period and its first harmonic are marked by a thin vertical line. $(p_1,p_2)$ and $(p^\prime_1,p^\prime_2)$ denote the first two strongest peaks near the orbital period, and their corresponding first harmonics, {respectively}. The vertical dashed lines show intervals with half-width $10^{-4}$~day around the half-periods. (b) LS periodogram applied to the simulated drifting spot clusters in Fig.~\ref{mig_rand}b. {The vertical lines from panel (a) are kept for comparison.} }
\label{fig:powerfull}
\end{figure}

We applied a Lomb-Scargle (LS) periodogram to the entire {model-subtracted} dataset, {$F^{\rm M}_{30}-F_i$, which allows} us to investigate the full power spectrum with the eclipse effects largely eliminated. Figure~\ref{fig:powerfull}a shows two parts from the 
LS periodogram, {with actual data and} with an oversampling factor of 10. There are several peaks in the vicinity of the orbital period, which can be ascribed to rotational modulation owing to spot groups with different angular velocities, possibly due to different latitudes. However, the actual period resolution is about $10^{-5}$~d, {leaving any} closely packed multiple peaks near the orbital {first} harmonic {unresolved}. In addition, there are a few peaks in the harmonic range, which result from frequency couplings in the orbital-period range. For instance, the central isolated peak near 0.2635~d is due to a modulation of $p_1$ with $p_{\rm orb}$. 

To estimate the type of {SDR}, we apply the procedure recently proposed by \citet{ra15} and further investigated by \citet{santos17}. 
{The method is based on the argument that lower latitude spots lead to less sine-shaped rotational modulations, which imparts more power into harmonics, hence increasing the power ratio of the first harmonic to the fundamental (rotational) period. Comparing periods of the highest peaks (near the stellar rotation period) and the related power ratios, the method thus aims to identify the sign of differential rotation \citep[see][for limitations and possible ambiguities]{santos17}.}
We identified two peaks with the highest power for the region near the orbital period (Fig.~\ref{fig:powerfull}a, {lower panel}), {$p_1,p_2$} and the comparison region near its first harmonic (half the orbital period; Fig.~\ref{fig:powerfull}a). The first harmonics of the two highest peaks, at $p^\prime_1$ and $p^\prime_2$, fall within $\pm 10^{-4}$ of the corresponding half-periods. This is far below the $\pm 10^{-2}$ detection threshold used for a large sample of single stars by \citet{ra15}. {Considering} the peaks $(p_1,p_2)$, we find primary to harmonic peak height ($h$) ratios {($r_k := h(p^\prime_k)/h(p_k)$)} of $r_1=0.039$, $r_2=0.024$. {This means that $p_1$ emanates from a spot which causes less sinusoidal rotational modulation and thus lies at a lower latitude than the spot responsible for $p_2$. Because $p_1<p_2$, we find \emph{solar-like differential rotation}} for the K1-type sub-giant component.\footnote{Upon our request, T. Reinhold (priv. comm.) has confirmed our result by kindly applying their \emph{generalised} LS periodogram procedure.} Combining this indication with our detection that the majority of drifts are in a prograde sense (Figs.~\ref{fig:map} and \ref{fig:histomed}), the scheme outlined in Table~\ref{tab:sync} implies that $\lambda_{\rm sync}$ is {a higher} latitude {compared to the mean latitude of spots which lead to rotational modulation (the second case in the list)}. 

The observed range [$p_1,p_2$] of rotation periods corresponds to a differential rotation of $2.77^\circ$/d, which is not far from the estimated range of $3.79^\circ$/d obtained in Sect.~\ref{sssec:drtime} using longitudinal shifts. Figure~\ref{fig:powerfull}b shows the same period ranges for the thousand-cycle simulation presented in Fig.~\ref{mig_rand}. The strongest peak {corresponds to about $1.9^\circ$/d (same as the input)} relative rotation rate of both spot clusters, {and it} is very close to the location of the {observed} $p_1$ in Figure~\ref{fig:powerfull}a. The peak on panel (b) {between $p_1^\prime$ and $p_2^\prime$}, is due to half the orbital period, and shows that the eclipse effects are not fully avoided, {even though} we subtract the model light curve. So we put a word of caution for the applicability of the method of \citet{ra15}, which the authors applied to single stars. 

Persistent spot activity in pairs of latitudinal bands in the same rotational hemisphere can lead to low-frequency beats, resembling dynamo-related cycles. To investigate this possibility, we calculated limits for the period of such a beat wave, when the frequency pairs $(p_1,p_2)$ {in Fig.~\ref{fig:powerfull}a} and a close pair ($p_2$ and its {short-period} neighbour) are taken at their face value. This results in a period interval of (0.7,3.0) years, which means that the peaks falling into this range in the LS periodograms in Fig.~\ref{fig:periodograms} could be caused by such an indirect effect of latitudinal differential rotation. 

\begin{figure}
\centering
    \includegraphics[width=\columnwidth]{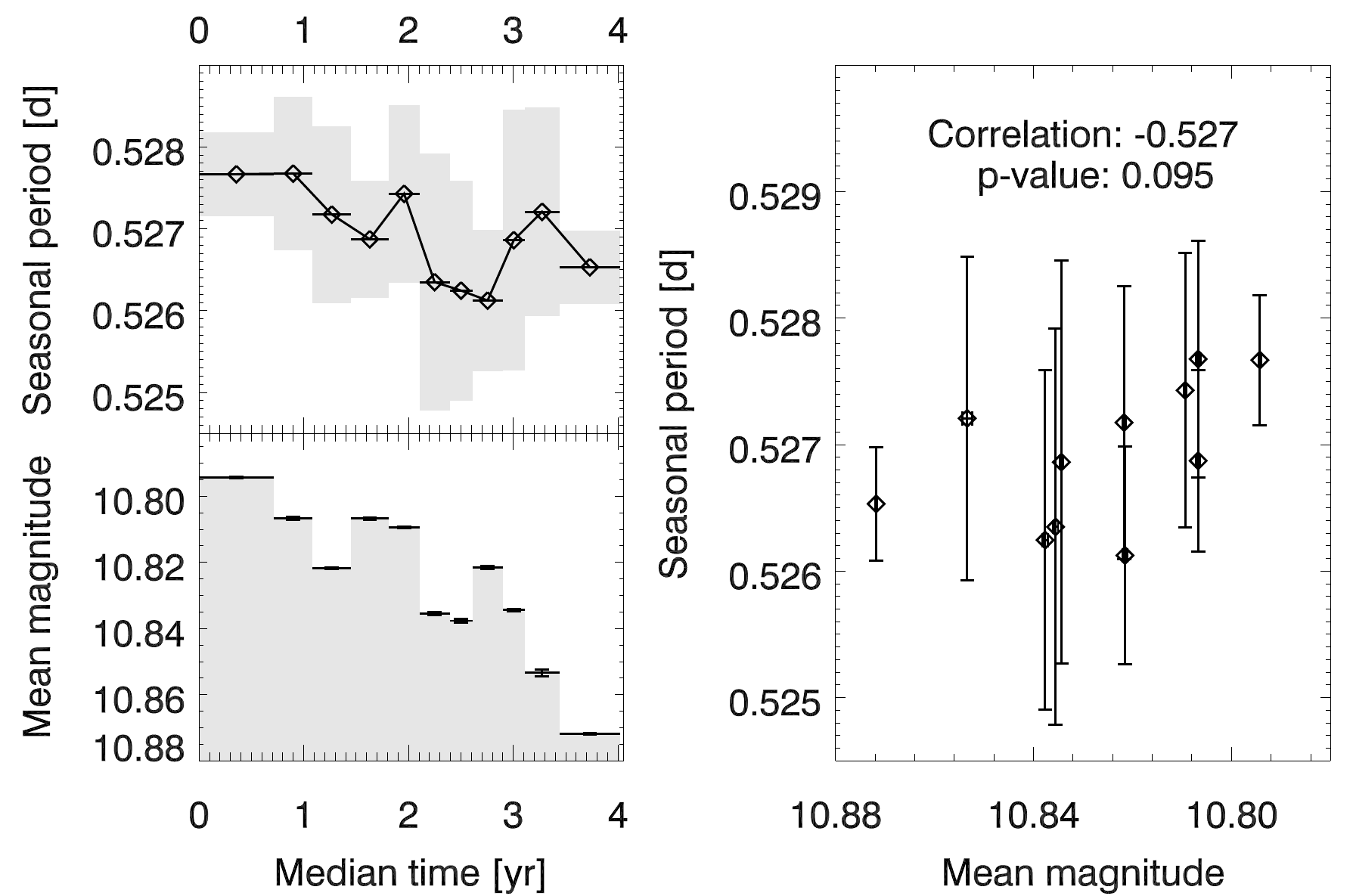}
    \caption{\emph{Left panel:} Time series of the seasonal periods {(upper plot). FWHM of the determined peaks in periodograms are shown with shaded vertical bands.} The mean system brightness (lower plot). The standard deviations of the means are of order $10^{-4}$ mag {(vertical lines)}. \emph{Right panel:} Seasonal {periods versus mean brightness}. The error bars show the {FWHMs of the periodogram peaks}.}
\label{fig:pers}
\end{figure}

To isolate the rotation periods of individual spot complexes visually detected in Fig.~\ref{long_spots}, we have generated LS periodograms for 11 seasons marked in Fig.~\ref{maxs_fs}, in a way similar to the partitioning done by \citet{ioan2016}. 
The selection of seasons is consistent with the emergence and disappearance of major spot regions in Fig.~\ref{fig:map}. We detected single peaks close to the orbital period for each season and noticed that 
they lie within the range of the multiple peaks near the orbital period in Fig.~\ref{fig:powerfull}b.  The left panel of Fig.~\ref{fig:pers} shows the time variation of the seasonal periods (upper figure) {and the seasonal means of total brightness (lower figure), showing similarity with the seasonal period variation. However, the FWHMs of the seasonal periodogram peaks (shades in the upper figure) are comparable with the period fluctuation in annual time scales, hampering further conclusions about a parallelism with the brightness variation. If this fluctuation is real, then} the period between {the three local} maxima is about 1.3 yr, which is about the same as the periodogram peak obtained in Sect.~\ref{ssec:cycles} from mean brightness variations. 

On the right panel of Fig.~\ref{fig:pers}, we present the seasonal averages of relative rotation rate (Sect.~\ref{sssec:drtime}) as a function of the corresponding seasonal period. We find a correlation of {$-0.53$ with a p-value of $0.10$. A possible physical relationship between overall darkening and the shortening in spot periods is the equatorward propagation of activity.}

\section{Discussion}
\label{sec:discuss}

We have presented a detailed investigation of activity variations and differential 
rotation in the K1-type sub-giant component of the close binary system KIC 11560447, based on practically uninterrupted \emph{Kepler} photometry. 
When focusing on spot patterns and their rotation on the K1IV star, we neglected 
the contribution of the M-type red dwarf component, which is responsible for only 3 percent of the total luminosity of this close binary system. 

\subsection{Spot activity}

Through numerical simulations, we have tested the performance of {\tt DoTS} in reconstructing input data. To avoid spurious `active longitudes' owing to systematic errors amplified by unresolved spots, it is essential to determine effective temperatures and radii of binary components with sufficient accuracy, prior to light curve inversions \citep{Jeffers2005}. {\tt DoTS} then becomes sensitive to large-scale longitudinal gradients in the spot filling factor. Even when the entire surface is peppered with small-scale spots, spontaneous clustering of random spots can be correctly detected, as we demonstrated in the current study, thanks to precise determination of system parameters. The two drifting spot clusters moving in the prograde direction on top of a randomly changing background of smaller spots (Fig.~\ref{mig_rand}) are reminiscent of the picture recently recalled by \citet{jarvinen15}, {for} the possible co-existence of short-living small spots \citep{mosser09} and the long-living large spot conglomerates \citep{hussain02}. 
Our study underlines the importance of forward modelling of surface rotation and activity, when interpreting reconstructions based on real data.

Inversions of light curves covering $\sim 2800$ orbital cycles and the flux contrast variation of KIC 11560447 have shown that the spot emergence pattern of the K1 sub-giant is predominantly nonaxisymmetric.
The separation of $3.59R_\odot$ between the components and the 0.53-day rotation period hints at tidal effects being responsible for the preferred longitudes. The effect was modelled by \citet{holzwarth03} by simulations of magnetic flux tubes erupting through a tidally distorted convection zone of a Sun-like stellar model. The authors found an increasing degree of longitudinal clustering for 1-day rotation period as compared to 2-day, and in some cases they found completely avoided longitude intervals. In Fig.~\ref{fig:map}, the spots detected around $\phi=300^\circ$ and $60^\circ$ are primarily those which migrate in longitude in the orbital frame. Though it is difficult to track the emergence longitudes of such spot groups, in a few cases the spots emerge near the active longitude, migrate prograde (presumably by differential rotation), and eventually pass through the `avoided' zone. 

In the course of the 4-year observing period, the system showed a secular decrease in total brightness ($\sim 0.1$~mag), which is predominantly led by the primary star, because the secondary component would have to fade by about 2 magnitudes to 
be responsible for the effect. This is much less likely than a long-term variation on the K1IV component. Likewise, the detected spot clusters (hence their drifts) are by far dominated by the K1IV component, {because} their average {flux} contrast is comparable with that of the secular change. 

The secular dimming occurs across all longitudes, while inversions still detect relatively strong, drifting spot regions (Fig.~\ref{fig:map}). In addition, there is evidence for an overall shortening of the mean spot rotation period (Fig.~\ref{fig:pers}). 
The underlying processes can be (1) the gradual formation of a sufficiently dark polar cap and/or (2) a sufficiently high spot filling factor developing at low-mid latitudes (at all longitudes), which can be driven by a solar-like activity cycle. 
There are signs of an activity cycle of about 1.3 year as seen in the system brightness and the primary maximum level, but this could also be a manifestation of prolonged {spot} emergence in half-year periods at {preferred} longitude zones.

\subsection{Differential rotation}

Thanks to continuous \emph{Kepler} monitoring for about 4 years, we were able to detect a range of rotational periods of major spot groups in KIC 11560447. Our detection is not limited to periodograms, but we independently tracked multiple spot patterns in longitude and time. We measured a range of slopes for spot groups presumably at different latitudes, rotating with different angular velocities. However, the signatures of surface differential rotation can be contaminated by near-surface radial shear and/or the anchoring depth of emerged magnetic features. We plan to carry out similar measurements in conjunction with other indicators on a set of active stars observed with \emph{Kepler}. This would provide comparative results and a means to constrain SDR in single and binary stars of different types and evolutionary stages.

We cannot make any conclusive statement about the strength of SDR, because we do not know the latitudinal range of spots which cause rotational modulation. In a rapidly rotating late-type sub-giant star, we theoretically expect a high-latitude confinement 
of spot activity, owing to strong Coriolis acceleration of rising flux tubes and large fractional thickness of the convection zone \citep{isik11}. Therefore, the accuracy of pole-equator lag estimates via photometric analysis in rotational timescale might be very low.

The periodogram of the model-subtracted data indicates solar-like SDR, using the method of \citet{ra15}. However, it should be noted that periodicities driven by the eclipsing binary geometry might still have survived in the model-subtracted light curve, casting some doubt on the applicability of this method in our case. Our result relies on the assumption that both peaks $p_1$ and $p_2$ purely stem from spot modulation.

The majority of spots drift in the prograde sense, ie., faster than the orbital rate. Solar-like SDR then implies that a relatively high-latitude zone rotates synchronously with the orbit. It should be noted that higher-latitude spots would yield weaker rotational modulations, so our inversions could also have underestimated their contributions.

\section{Conclusions}
\begin{itemize}
\item {Our procedure using the {\tt DoTS} code can detect spontaneous occurrence of clusters of unresolved spots. It can also track up to 3-4 migrating spot groups which are well-separated and migrating in longitude. In both cases, spurious longitudes are avoided when precise stellar parameters are available, assuming a uniform probability distribution of unresolved background spots across the surface. }
\item {Recurrent} starspot activity integrated over 4 years of {\emph{Kepler} photometry of KIC 11560447} appears to cluster {in} two longitudinal zones centred around $0^\circ$ and $140^\circ$. The flux contrast PDF {supports} that these {can be} mildly preferred starspot emergence longitudes, {while} the subsequent evolution of activity {continues} at all orbital longitudes. 
\item The rotation of active regions {is} largely prograde in the orbital frame ({up to about} $2.4^\circ$/d), and there is an indication for solar-like differential rotation. However, as the relative motion occurs {is detected} in such long-lived active nests, the drifts can also be {contributed by} the {continuous magnetic flux emergence} with a systematic prograde component, or azimuthal dynamo waves. 
\item {In case of solar-like SDR, relatively high-latitudes must be rotating near-}synchronously with the orbit, {to explain the majority of spots drifting prograde, presumably at lower latitudes. }
\item The axisymmetric component of the spot coverage  increases substantially towards the end of the observation period, while the mean spot period decreases. In case of solar-like SDR, this is indicative of equatorward propagation of activity and/or {strong} circumpolar accumulation of {dark spots, which is expected in rapidly rotating RS~CVn systems}. 
\end{itemize}

Long-term multi-colour photometric and spectropolarimetric monitoring 
of this binary system, as well as studies devoted to forward modelling of starspots in the presence of surface shear would be very useful in the quest of characterising activity and differential rotation in rapidly rotating active cool stars. 

\section*{Acknowledgements}

{The authors are grateful to the anonymous referee for careful and critical comments, which helped substantially to improve the manuscript.}
H.V. \c{S}enavc{\i} acknowledges the support by The Scientific and Technological Research Council of Turkey (T\"{U}B\.{I}TAK) through the project grant 1001 - 115F033. E. I\c{s}\i k acknowledges support by the Young Scientist Award Programme BAGEP-2016 of the Science Academy, Turkey, and thankful to T. Reinhold and R.H. Cameron for useful discussions. GAJH would also like to acknowledge the IDEX initiative at Universit\'{e} F\'{e}d\'{e}rale Toulouse Midi-Pyr\'{e}n\'{e}es (UFTMiP) for funding the STEPS programme enabling regular research visits. This work has made use of the VALD database, operated at Uppsala University, the Institute of Astronomy RAS in Moscow, and the University of Vienna.




\bibliographystyle{mnras}
\bibliography{kic11560447.bib}




\appendix

\section{Longitudinal drift rates}
In Table~\ref{table3} we list the results from the linear fitting procedure applied in Sect.~\ref{sssec:drtime}. The columns denote, for each drifting feature in Fig.~\ref{long_spots} the index number $i$, the median time $t_i$ in years since BJD0, the lifetime $\Delta t_i$ in years, the median longitude $\phi_i$ in degrees, the longitudinal range $\Delta\phi_i$ in degrees, the rotation rate relative to the orbital frame $\Delta\phi_i$ in degrees per day that results from the linear fit, and the standard deviation $\sigma_i$, in longitudinal degrees, of the filtered data points about the fitting line. Note that the fits were made to median longitudes. 
\begin{table}
\caption{The results of the median longitude fitting procedure for LC and SC data.}
\label{table3}
\centering
Long cadence (LC)
\begin{tabular}{p{5mm}rrrrrr}
\hline
\emph{i}& t$_i$(yr) & $\Delta t_i$(d) & $\phi_i(^\circ)$ & $\Delta\phi_i(^\circ)$ & $\Delta\Omega_i(^\circ/d)$ & $\sigma(^\circ)$ \\
\hline
0	&	0.05	&	32.7	&	364.0	&	72.4	&	1.39	&	17.2	\\
7	&	0.21	&	52.8	&	132.7	&	60.3	&	0.42	&	12.7	\\
1	&	0.24	&	41.2	&	370.1	&	68.4	&	0.50	&	13.8	\\
2	&	0.34	&	24.3	&	363.0	&	66.4	&	-0.14	&	14.9	\\
8	&	0.36	&	46.4	&	109.6	&	50.3	&	-0.58	&	11.9	\\
3	&	0.47	&	57.0	&	378.1	&	100.6	&	0.59	&	24.4	\\
4	&	0.60	&	26.9	&	379.1	&	54.3	&	0.03	&	13.9	\\
9	&	0.62	&	65.4	&	124.7	&	100.6	&	1.25	&	26.0	\\
5	&	0.73	&	21.1	&	330.8	&	50.3	&	-1.35	&	11.9	\\
6	&	0.82	&	39.0	&	317.8	&	64.4	&	0.06	&	13.4	\\
10	&	0.84	&	51.2	&	203.1	&	64.4	&	0.45	&	14.6	\\
11	&	1.09	&	32.7	&	31.2	&	54.3	&	1.16	&	12.8	\\
12	&	1.18	&	74.9	&	90.5	&	44.2	&	0.50	&	12.8	\\
13	&	1.29	&	67.5	&	180.0	&	102.6	&	0.28	&	26.1	\\
14	&	1.55	&	50.7	&	383.1	&	50.3	&	0.42	&	10.7	\\
15	&	1.71	&	22.2	&	343.9	&	52.3	&	-0.74	&	12.6	\\
18	&	1.81	&	37.5	&	154.9	&	112.6	&	2.35	&	27.6	\\
16	&	1.82	&	33.8	&	337.9	&	52.3	&	0.05	&	12.4	\\
19	&	1.91	&	22.2	&	201.1	&	36.2	&	0.32	&	7.6	\\
20	&	1.99	&	22.7	&	266.5	&	54.3	&	0.46	&	11.3	\\
17	&	2.01	&	91.8	&	359.0	&	98.5	&	0.71	&	23.5	\\
21	&	2.06	&	21.1	&	264.5	&	50.3	&	1.09	&	11.3	\\
22	&	2.31	&	99.7	&	153.9	&	86.5	&	0.48	&	17.5	\\
23	&	2.53	&	44.3	&	258.4	&	102.6	&	2.44	&	28.2	\\
25	&	2.65	&	15.8	&	125.7	&	30.2	&	-0.30	&	6.5	\\
24	&	2.67	&	45.9	&	365.0	&	86.5	&	0.94	&	19.2	\\
26	&	2.72	&	22.2	&	142.8	&	60.3	&	1.47	&	12.7	\\
27	&	2.83	&	45.9	&	160.9	&	100.6	&	0.57	&	24.7	\\
28	&	3.00	&	69.1	&	198.1	&	110.6	&	0.98	&	22.9	\\
29	&	3.11	&	63.3	&	24.1	&	64.4	&	0.37	&	11.3	\\
31	&	3.33	&	131.4	&	220.2	&	114.6	&	0.57	&	24.0	\\
30	&	3.36	&	100.8	&	102.6	&	160.9	&	1.04	&	32.9	\\
32	&	3.61	&	77.0	&	387.2	&	54.3	&	0.11	&	10.2	\\
35	&	3.63	&	26.4	&	256.4	&	58.3	&	1.37	&	10.9	\\
33	&	3.80	&	35.4	&	387.2	&	58.3	&	0.29	&	12.0	\\
34	&	3.93	&	48.5	&	379.1	&	42.2	&	-0.42	&	11.2	\\
36	&	3.96	&	50.1	&	156.9	&	100.6	&	1.42	&	21.6	\\

\hline
\multicolumn{7}{|c|}{Short cadence (SC)}\\
\hline
0	&	0.07	&	25.3	&	14.1	&	76.4	&	1.23	&	15.8	\\
1	&	0.24	&	18.5	&	13.1	&	50.3	&	0.34	&	12.7	\\
2	&	0.26	&	28.5	&	139.8	&	50.3	&	0.23	&	11.2	\\
7	&	2.69	&	7.90	&	126.7	&	24.1	&	-0.65	&	5.4	\\
3	&	2.70	&	16.4	&	27.2	&	42.2	&	0.45	&	9.5	\\
8	&	2.74	&	13.7	&	147.8	&	54.3	&	1.88	&	12.8	\\
9	&	2.80	&	32.7	&	166.9	&	120.7	&	1.03	&	29.9	\\
10	&	2.88	&	14.8	&	160.9	&	64.4	&	0.46	&	16.4	\\
11	&	2.89	&	9.5	&	248.4	&	34.2	&	1.08	&	7.2	\\
5	&	3.01	&	7.4	&	202.1	&	50.3	&	1.52	&	12.9	\\
4	&	3.05	&	17.9	&	13.1	&	46.3	&	0.06	&	12.2	\\
6	&	3.06	&	8.4	&	221.2	&	44.3	&	0.28	&	11.2	\\

\hline
\end{tabular}
\end{table}





\bsp	
\label{lastpage}
\end{document}